\documentclass[aps,preprint]{revtex4}%
\usepackage{amsfonts}
\usepackage{amsmath}
\usepackage{amssymb}
\usepackage{subfigure}
\usepackage{graphicx}%
\begin{document}
\preprint{CTP-SCU/2015010}
\title{Thermodynamics and Luminosities of Rainbow Black Holes}
\author{Benrong Mu$^{a,b}$}
\email{mubenrong@uestc.edu.cn}
\author{Peng Wang$^{b}$}
\email{pengw@scu.edu.cn}
\author{Haitang Yang$^{b,c}$}
\email{hyanga@scu.edu.cn}
\affiliation{$^{a}$Physics Teaching and Research Section, College of Medical Technology,
Chengdu University of Traditional Chinese Medicine, Chengdu 611137, China}
\affiliation{$^{b}$Center for Theoretical Physics, College of Physical Science and
Technology, Sichuan University, Chengdu, 610064, China}
\affiliation{$^{c}$Kavli Institute for Theoretical Physics China (KITPC), Chinese Academy of Sciences, Beijing 100080, China}

\begin{abstract}
Doubly special relativity (DSR) is an effective model for encoding quantum
gravity in flat spacetime. As a result of the nonlinearity of the Lorentz
transformation, the energy-momentum dispersion relation is modified. One
simple way to import DSR to curved spacetime is \textquotedblleft Gravity's
rainbow", where the spacetime background felt by a test particle would depend
on its energy. Focusing on the \textquotedblleft Amelino-Camelia dispersion
relation" which is $E^{2}=m^{2}+p^{2}\left[  1-\eta\left(  E/m_{p}\right)
^{n}\right]  $ with $n>0$, we investigate the thermodynamical properties of a
Schwarzschild black hole and a static uncharged black string for all possible
values of $\eta$ and $n$ in the framework of rainbow gravity. It shows that
there are non-vanishing minimum masses for these two black holes in the cases
with $\eta<0$ and $n\geq2$. Considering effects of rainbow gravity on both the
Hawking temperature and radius of the event horizon, we use the geometric
optics approximation to compute luminosities of a 2D black hole, a
Schwarzschild one and a static uncharged black string. It is found that the
luminosities can be significantly suppressed or boosted depending on the
values of $\eta$ and $n$.

\end{abstract}
\maketitle
\tableofcontents



\section{Introduction}

The Hawking radiation was a remarkable prediction of quantum field theory in
curved spacetime. As Stephen Hawking demonstrated\cite{IN-Hawking:1974sw}, a
Schwarzschild black hole emits radiation just like an ordinary blackbody at
temperature $T=\hbar\kappa/2k_{B}\pi,$ where $\kappa$ is the surface gravity.
Soon after this discovery, it was realized that there might be the
"trans-Planckian" problem\cite{IN-Unruh:1976db}. It appears that the Hawking
radiation originates from the modes with huge initial frequencies, well beyond
the Planck mass $m_{p}$, which undergo exponential high gravitational
red-shifting near the horizon. As a result, the Hawking's prediction relies on
the validity of quantum field theory in curved spacetime to arbitrary high
energies. On the other hand, quantum field theory is considered more like an
effective field theory of an underlying theory whose nature remains unknown.
This observation poses the question of whether any unknown physics at the
Planck scale could strongly influence the Hawking radiation.

Although a complete understanding of the trans-Planckian problem requires a
full theory of quantum gravity, there are various attempts using effective
models to address this problem. Among them is Doubly Special Relativity
(DSR)\cite{IN-AmelinoCamelia:2000ge,IN-AmelinoCamelia:2000mn,IN-Magueijo:2001cr,IN-Magueijo:2002am}%
, where the transformation laws of special relativity are modified at very
high energies. In DSR, the energy-momentum dispersion relation for a massive
particle of mass $m$ is modified to%
\begin{equation}
E^{2}f^{2}\left(  E/m_{p}\right)  -p^{2}g^{2}\left(  E/m_{p}\right)  =m^{2},
\label{eq:MDR}%
\end{equation}
where $m_{p}$ is the Planck mass, and $f\left(  x\right)  $ and $g\left(
x\right)  $ are two unknown functions with the following properties%
\begin{equation}
\lim_{x\rightarrow0}f\left(  x\right)  =1\text{ and }\lim_{x\rightarrow
0}g\left(  x\right)  =1\text{.}%
\end{equation}
It has been shown that the modified dispersion relation (MDR) might play a
role in astronomical and cosmological observations, such as the threshold
anomalies of ultra high energy cosmic rays and TeV
photons\cite{IN-AmelinoCamelia:1997gz,IN-Colladay:1998fq,IN-Coleman:1998ti,IN-AmelinoCamelia:2000zs,IN-Jacobson:2001tu,IN-Jacobson:2003bn}%
. Moreover, thermodynamics of black holes have been explored in the framework
of the
MDR\cite{IN-AmelinoCamelia:2004xx,IN-Ling:2005bq,IN-AmelinoCamelia:2005ik,IN-Nozari:2006ka,IN-Sefiedgar:2010we,IN-Majumder:2011xg}%
.

One of the most popular choice for the functions $f\left(  x\right)  $ and
$g\left(  x\right)  $ has been proposed by Amelino-Camelia et
al.\cite{IN-AmelinoCamelia:1996pj,IN-AmelinoCamelia:2008qg}, which gives%
\begin{equation}
f\left(  x\right)  =1\text{ and }g\left(  x\right)  =\sqrt{1-\eta x^{n}}.
\label{eq:AC-Dispersion}%
\end{equation}
Usually one has $n>0$. As shown in \cite{IN-AmelinoCamelia:2008qg}, this
formula is compatible with some of the results obtained in the
Loop-Quantum-Gravity approach and reflects the results obtained in $\kappa
$-Minkowski and other noncommutative spacetimes. Phenomenological implications
of this \textquotedblleft Amelino-Camelia (AC) dispersion relation" are also
reviewed in \cite{IN-AmelinoCamelia:2008qg}.

Nevertheless, the non-linear realization of the Lorentz transformation in the
framework of DSR results in a very complicated definition of the dual position
space. To circumvent this difficulty, Magueijo and
Smolin\cite{IN-Magueijo:2002xx} proposed the \textquotedblleft Gravity's
rainbow", where the spacetime background felt by a test particle would depend
on its energy. Consequently, the energy of the test particle deforms the
background geometry and hence the dispersion relation. As regards the metric,
it would be replaced by a one parameter family of metrics which depends on the
energy of the test particle, forming a \textquotedblleft rainbow\ metric".
Specifically, for the Schwarzschild solution, the corresponding
\textquotedblleft rainbow metric\textquotedblright\ is
\begin{equation}
ds^{2}=\left(  1-\frac{2GM}{r}\right)  \frac{dt^{2}}{f^{2}\left(
E/m_{p}\right)  }-\frac{1}{g^{2}\left(  E/m_{p}\right)  }\left[  \frac{dr^{2}%
}{1-\frac{2GM}{r}}+r^{2}\left(  d\theta^{2}+\sin^{2}\theta d\phi^{2}\right)
\right]  , \label{eq:Rainbow-Schwarzchild-Metric}%
\end{equation}
which is the spherically symmetric solution to the distorted Einstein's Field
equations given in \cite{IN-Magueijo:2002xx}. Note that $E$ is the energy of
the probing particle. To obtain the modified Schwarzschild metric from the
usual one, it appears from eqn. $\left(  \ref{eq:Rainbow-Schwarzchild-Metric}%
\right)  $ that we can simply make replacements $dt\rightarrow dt/f\left(
E/m_{p}\right)  $ for the time coordinate and $dx^{i}\rightarrow
dx^{i}/g\left(  E/m_{p}\right)  $ for all spatial coordinates. In fact, such
procedure also works for the black objects besides the Schwarzschild
metric\cite{IN-Ali:2014zea}. The rainbow gravity formalism has received a lot
of attentions recently, for instance, in
cosmology\cite{In-Ling:2006az,IN-Ling:2008sy,IN-Garattini:2012ca,IN-Awad:2013nxa}
and black hole
physics\cite{IN-Ling:2005bp,In-Galan:2006by,IN-Li:2008gs,IN-Garattini:2009nq,IN-Salesi:2009kd,IN-Esposito:2010pg,IN-Ali:2014xqa,IN-Gim:2014ira}%
.

In this paper, we will study thermodynamics and luminosities of black holes in
the framework of rainbow gravity. The remainder of our paper is organized as
follows. In section \ref{Sec:HJM}, the deformed Hamilton-Jacobi equations for
scalars, spin $1/2$ fermions and vector bosons are derived in the framework of
rainbow gravity. We then solve the Hamilton-Jacobi equations to obtain
tunneling rates. The temperatures and entropies of a rainbow Schwarzschild
black hole and a rainbow static uncharged black string is computed in section
\ref{Sec:TBHRG}. In section \ref{Sec:LBHRG}, we calculate the luminosities of
a 2D rainbow black hole, a 4D rainbow spherically symmetric one, and a 4D
rainbow cylindrically symmetric one. Section \ref{Sec:Con} is devoted to our
conclusion. Throughout the paper we take Geometrized units $c=G=1$, where the
Planck constant $\hbar$ is square of the Planck mass $m_{p}$.

\section{Hamilton-Jacobi Method}

\label{Sec:HJM}

After the Hawking's original derivation, there have been some other methods
proposed to understand the Hawking radiation. Recently, a semiclassical method
of modeling Hawking radiation as a tunneling process has been developed and
attracted a lot of attention. This method was first proposed by Kraus and
Wilczek\cite{HJM-Kraus:1994by,HJM-Kraus:1994fj}, which is known as the null
geodesic method. They employed the dynamical geometry approach to calculate
the imaginary part of the action for the tunneling process of s-wave emission
across the horizon and related it to the Hawking temperature. Later, the
tunneling behaviors of particles were investigated using the Hamilton-Jacobi
method\cite{HJM-Srinivasan:1998ty,HJM-Angheben:2005rm,HJM-Kerner:2006vu}. In
the Hamilton-Jacobi method, one ignores the self-gravitation of emitted
particles and assumes that its action satisfies the relativistic
Hamilton-Jacobi equation. The tunneling probability for the classically
forbidden trajectory from inside to outside the horizon is obtained by using
the Hamilton-Jacobi equation to calculate the imaginary part of the action for
the tunneling process.

In this section, the Hamilton-Jacobi equations for scalars, spin $1/2$
fermions and vector bosons in the rainbow metric are derived. The particles'
tunneling rates across the event horizon $r=r_{h}$ of the rainbow metric
$\left(  \ref{eq:Rainbow-Metric}\right)  $ are then computed by solving the
Hamilton-Jacobi equations. We here consider a static black hole with the line
element%
\begin{equation}
ds^{2}=B\left(  r\right)  dt^{2}-\frac{dr^{2}}{B\left(  r\right)  }-C\left(
r^{2}\right)  h_{ab}\left(  x\right)  dx^{a}dx^{b}, \label{eq:BHmetric}%
\end{equation}
whose rainbow metric is obtained by $dt\rightarrow dt/f\left(  E/m_{p}\right)
,dr\rightarrow dr/g\left(  E/m_{p}\right)  $, and $dx_{i}\rightarrow
dx_{i}/g\left(  E/m_{p}\right)  $
\begin{equation}
ds^{2}=\frac{B\left(  r\right)  dt^{2}}{f^{2}\left(  E/m_{p}\right)  }%
-\frac{dr^{2}}{g^{2}\left(  E/m_{p}\right)  B\left(  r\right)  }%
-\frac{C\left(  r^{2}\right)  h_{ab}\left(  x\right)  dx^{a}dx^{b}}%
{g^{2}\left(  E/m_{p}\right)  }. \label{eq:Rainbow-Metric}%
\end{equation}
The function $B\left(  r\right)  $ has a simple zero at $r=r_{h}$ with
$B^{\prime}\left(  r_{h}\right)  $ being finite and nonzero. The vanishing of
$B\left(  r\right)  $ at point $r=r_{h}$ indicates the presence of an event
horizon. For simplicity, we assume that the particles are massless.

In the rainbow metric $ds^{2}=\tilde{g}_{\mu\nu}\left(  E\right)  dx^{\mu
}dx^{\nu}$, the massless scalar field $\phi$ obeys the Klein-Gordon equation%
\begin{equation}
\tilde{\nabla}^{\mu}\tilde{\nabla}_{\mu}\phi=0, \label{eq:KG}%
\end{equation}
where $\tilde{\nabla}_{\mu}$ is the covariant derivative associated with
$\tilde{g}_{\mu\nu}\left(  E\right)  $ and the index $\mu$ is lowered or
raised by $\tilde{g}_{\mu\nu}\left(  E\right)  $. Making the ansatz for $\phi$
which is%
\begin{equation}
\phi=\exp\left(  \frac{iI}{\hbar}\right)  ,
\end{equation}
substituting it into eqn. $\left(  \ref{eq:KG}\right)  $, and expanding eqn.
$\left(  \ref{eq:KG}\right)  $ in powers of $\hbar$, the leading order gives
the Hamilton-Jacobi equation for a massless scalar particle
\begin{equation}
\tilde{g}_{\mu\nu}\left(  E\right)  \partial^{\mu}I\partial^{\nu}I=0.
\label{eq:Hamilton-Jacobi}%
\end{equation}

The Dirac equation for a spin-$1/2$ fermion field $\psi$ takes the form of%
\begin{equation}
i\tilde{\gamma}_{\mu}\left(  \partial^{\mu}+\tilde{\Omega}^{\mu}\right)
\psi=0, \label{eq:Dirac}%
\end{equation}
where $\tilde{\Omega}_{\mu}\equiv\frac{i}{2}\tilde{\omega}_{\mu}^{\text{ }%
ab}\tilde{\Sigma}_{ab}$, $\tilde{\Sigma}_{ab}$ and $\tilde{\omega}_{\mu
}^{\text{ }ab}$ are the Lorentz spinor generator and spin connection in the
rainbow metric, respectively, and $\left\{  \tilde{\gamma}_{\mu},\tilde
{\gamma}_{\nu}\right\}  =2\tilde{g}_{\mu\nu}\left(  E\right)  $. To obtain the
Hamilton-Jacobi equation for the fermion, the ansatz for $\psi$ is assumed as
\begin{equation}
\psi=\exp\left(  \frac{iI}{\hbar}\right)  v, \label{eq:fermionansatz}%
\end{equation}
where $v$ is a slowly varying spinor amplitude. Substituting eqn. $\left(
\ref{eq:fermionansatz}\right)  $ into eqn. $\left(  \ref{eq:Dirac}\right)  $,
we find to the lowest order of $\hbar$%
\begin{equation}
\tilde{\gamma}_{\mu}\partial^{\mu}Iv=0. \label{eq:Hamilton-JacobiF}%
\end{equation}
Multiplying both sides of eqn. $\left(  \ref{eq:Hamilton-JacobiF}\right)  $
from the left by $\tilde{\gamma}_{\nu}\partial^{\nu}I$, one gets%
\begin{equation}
\left[  \tilde{g}_{\mu\nu}\left(  E\right)  \partial^{\mu}I\partial^{\nu
}I\right]  v=0.
\end{equation}
Since $v$ is nonzero, the Hamilton-Jacobi equation for a massless fermionic
particle is also given by eqn. $\left(  \ref{eq:Hamilton-Jacobi}\right)  $.

The Maxwell's equations for a massless vector field $A_{\mu}$ is%
\begin{equation}
\tilde{\nabla}^{\mu}\tilde{F}_{\mu\nu}=0, \label{eq:Maxwell-Equation}%
\end{equation}
where $\tilde{F}_{\mu\nu}=\tilde{\nabla}_{\mu}A_{\nu}-\tilde{\nabla}_{\nu
}A_{\mu}$. We then make the WKB ansatz%
\begin{equation}
A_{\mu}=a_{\mu}\exp\left(  \frac{iI}{\hbar}\right)  ,
\end{equation}
where $a_{\mu}$ is the polarization vector and $I$ is the action. Plugging the
WKB ansatz into eqn. $\left(  \ref{eq:Maxwell-Equation}\right)  $, we find
that leading order of $\hbar$ gives%
\begin{equation}
\tilde{g}^{\nu\sigma}\left(  E\right)  \left(  a_{\nu}\partial_{\sigma
}I\partial_{\mu}I-a_{\mu}\partial_{\sigma}I\partial_{\nu}I\right)  =0.
\label{eq:Hamilton-JacobiV}%
\end{equation}
To simplify eqn. $\left(  \ref{eq:Hamilton-JacobiV}\right)  $, one could
impose the Lorentz gauge in the curved spacetime%
\begin{equation}
\tilde{\nabla}^{\mu}A_{\mu}=0,
\end{equation}
where the leading order is
\begin{equation}
\tilde{g}^{\nu\sigma}\left(  E\right)  a_{\nu}\partial_{\sigma}I=0.
\label{eq:Lorentz-Gauge}%
\end{equation}
By plugging eqn. $\left(  \ref{eq:Lorentz-Gauge}\right)  $ into eqn. $\left(
\ref{eq:Hamilton-JacobiV}\right)  $, it shows that the Hamilton-Jacobi
equation for a massless vector boson is also given by eqn. $\left(
\ref{eq:Hamilton-Jacobi}\right)  $.

From eqn. $\left(  \ref{eq:Hamilton-Jacobi}\right)  ,$ one finds that the
Hamilton-Jacobi equation for a massless particle in the rainbow metric
$\left(  \ref{eq:Rainbow-Metric}\right)  $ becomes%
\begin{equation}
f^{2}\left(  E/m_{p}\right)  \frac{\left(  \partial_{t}I\right)  ^{2}%
}{B\left(  r\right)  }=g^{2}\left(  E/m_{p}\right)  \left[  B\left(  r\right)
\left(  \partial_{r}I\right)  ^{2}+\frac{h^{ab}\left(  x\right)  \partial
_{a}I\partial_{b}I}{C\left(  r^{2}\right)  }\right]  . \label{eq:HJ-Rainbow}%
\end{equation}
To solve the Hamilton-Jacobi equation for the action $I$, we can employ the
following ansatz%
\begin{equation}
I=-Et+W\left(  r\right)  +\Theta\left(  x\right)  ,
\end{equation}
where $E$ is the particle's energy. Plugging the ansatz into eqn. $\left(
\ref{eq:HJ-Rainbow}\right)  $, we have differential equations for $W\left(
r\right)  $ and $\Theta\left(  x\right)  $%
\begin{gather}
h^{ab}\left(  x\right)  \partial_{a}\Theta\left(  x\right)  \partial_{b}%
\Theta\left(  x\right)  =\lambda,\nonumber\\
\partial_{r}W_{\pm}\left(  r\right)  \equiv p_{r}^{\pm}=\frac{\pm1}{B\left(
r\right)  }\sqrt{\frac{E^{2}f^{2}\left(  E/m_{p}\right)  }{g^{2}\left(
E/m_{p}\right)  }-\lambda\frac{B\left(  r\right)  }{C\left(  r^{2}\right)  }},
\label{eq:HJ-Lamda&W}%
\end{gather}
where +/$-$ denotes the outgoing/ingoing solutions and $\lambda$ is a
constant. Using the residue theory for the semi circle, we get%
\begin{equation}
\operatorname{Im}W_{\pm}\left(  r\right)  =\frac{\pm\pi}{2\kappa}%
\frac{Ef\left(  E/m_{p}\right)  }{g\left(  E/m_{p}\right)  },
\end{equation}
where $\kappa=\frac{B^{\prime}\left(  r_{h}\right)  }{2}$. As shown in
\cite{SF}, the probability of a particle tunneling from inside to outside the
horizon is%
\begin{equation}
P_{emit}\propto\exp\left[  -\frac{2}{\hbar}\left(  \operatorname{Im}%
W_{+}-\operatorname{Im}W_{-}\right)  \right]  .
\end{equation}
There is a Boltzmann factor in $P_{emit}$ with an effective temperature, which
is
\begin{equation}
T_{eff}=T_{0}\frac{g\left(  E/m_{p}\right)  }{f\left(  E/m_{p}\right)  },
\label{eq:Eff-Temp}%
\end{equation}
where we define $T_{0}=\frac{\hbar\kappa}{2\pi}$ and take $k_{B}=1$.

\section{Thermodynamics of Black Holes in Rainbow Gravity}

\label{Sec:TBHRG}

We now use the Heisenberg uncertainty principle to estimate the black hole's
temperature. For a massless particle, the modified dispersion relation
$\left(  \ref{eq:MDR}\right)  $ becomes%
\begin{equation}
\frac{xf\left(  x\right)  }{g\left(  x\right)  }=y, \label{eq:MDR-Massless}%
\end{equation}
where $x=E/m_{p}\geq0$ and $y=p/m_{p}\geq0$. To obtain the black hole's
temperature from eqn. $\left(  \ref{eq:Eff-Temp}\right)  ,$ eqn. $\left(
\ref{eq:MDR-Massless}\right)  $ is needed to be solved for $x$ in terms of
$y$. In fact, one could have for $x$%
\begin{equation}
x=yh\left(  y\right)  , \label{eq:hFunction}%
\end{equation}
where eqn. $\left(  \ref{eq:MDR-Massless}\right)  $ is inverted to get the
function $h\left(  y\right)  $ and $\lim\limits_{y\rightarrow0}h\left(
y\right)  =1$. The Heisenberg uncertainty principle gives a relation between
the momentum $p$ of an emitted particle and the event horizon radius $r_{h}$
of the black hole\cite{TBHRG-Bekenstein:1973ur,TBHRG-Adler:2001vs}%
\begin{equation}
y=p/m_{p}\sim\delta p/m_{p}\sim\hbar/m_{p}\delta x\sim m_{p}/r_{h}.
\label{eq:HUP}%
\end{equation}
Substituting eqn. $\left(  \ref{eq:HUP}\right)  $ into eqn. $\left(
\ref{eq:Eff-Temp}\right)  $, we have for the black hole's temperature%
\begin{equation}
T=T_{0}\frac{E}{p}=T_{0}h\left(  \frac{m_{p}}{r_{h}}\right)  .
\label{eq:BH-Temp}%
\end{equation}
The range of the left-hand side(LHS) of eqn. $\left(  \ref{eq:MDR-Massless}%
\right)  $ determines the ranges of the values\ of $r_{h}$. Specifically, the
maximum value of the LHS of eqn. $\left(  \ref{eq:MDR-Massless}\right)  $,
which is denoted by $y_{cr}$, gives that $r_{h}\geq\frac{m_{p}}{y_{cr}}$. If
$y_{cr}$ is finite, it always predicts the existence of the black hole's
remnant. For some functions $f\left(  x\right)  $ and $g\left(  x\right)  $,
the domain of the LHS of eqn. $\left(  \ref{eq:MDR-Massless}\right)  $ might
be $\left[  0,x_{cr}\right]  $/$\left[  0,x_{cr}\right)  $ with $x_{cr}$ being
finite rather than $\left[  0,\infty\right)  $. Thus, it gives that the energy
of the particle $E\leq m_{p}x_{cr}$. If the domain is $\left[  0,\infty
\right)  $, we simply set $x_{cr}=\infty$.

\begin{figure}[tb]
\begin{centering}
\includegraphics[scale=0.9]{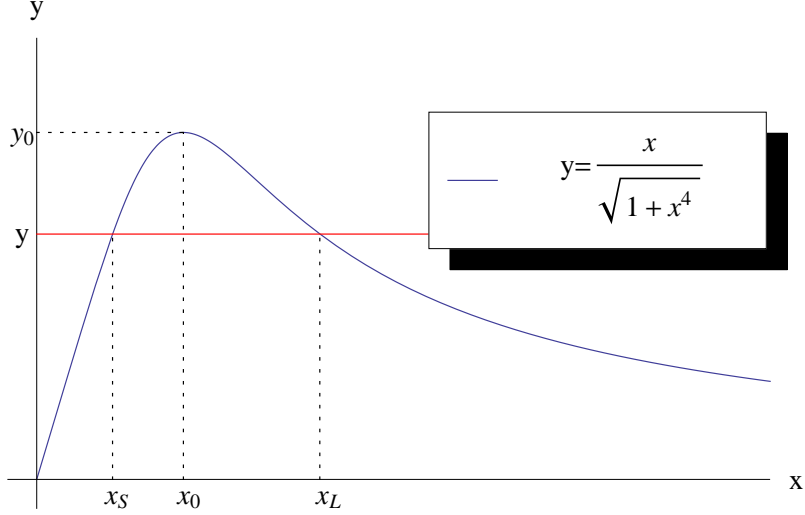}
\par\end{centering}
\caption{Plot of the curve of$\frac{x}{\sqrt{1-\eta x^{n}}}$ with $\eta=-1$
and $n=4$. The curve has a maximum value $y_{0}=\frac{1}{\sqrt{2}}$at
$x_{0}=1$. For $y<y_{0}$, there are $x_{S}$ and $x_{L}$ with $x_{S}<x_{L}$
satisfying $\frac{x_{L/S}}{\sqrt{1+x_{L/S}^{4}}}=y$. The solution $x_{L}$ is
considered runaways since it does not exist in the limit of $\eta\rightarrow
0$. Thus, only the solution $x_{S}$ is considered in the paper. }%
\label{fig:xvsy}%
\end{figure}

For the AC dispersion relation given in eqn. $\left(  \ref{eq:AC-Dispersion}%
\right)  $, eqn. $\left(  \ref{eq:MDR-Massless}\right)  $ becomes%
\begin{equation}
\frac{x}{\sqrt{1-\eta x^{n}}}=y. \label{eq:x&y-AC}%
\end{equation}
If $\eta>0,$ we find that $y_{cr}=0$. However, there is an upper bound
$x_{cr}=\eta^{-1/n}$ on $x$ to make the LHS of eqn. $\left(  \ref{eq:x&y-AC}%
\right)  $ real. If $\eta<0$, we have $x_{cr}=\infty$ and $y_{cr}=\infty$ for
$0<n<2$ and $x_{cr}=\infty$ and $y_{cr}=\left(  -\eta\right)  ^{-1/2}$ for
$n=2$. For the case with $\eta<0$ and $n>2,$ the LHS of eqn. $\left(
\ref{eq:x&y-AC}\right)  $ has a global maximum value $y_{0}$ at $x_{0}$, where
we define%
\begin{align}
x_{0}  &  =\left(  \frac{2-n}{2}\eta\right)  ^{-1/n},\nonumber\\
y_{0}  &  =\sqrt{\frac{n-2}{n}}\left(  \frac{2-n}{2}\eta\right)  ^{-1/n}.
\end{align}
Thus, it would appear that $y\leq y_{0}$ and $x<\infty$ since $x$ can go to
infinity. In FIG. $\ref{fig:xvsy}$, we plot an example with $\eta=-1$ and
$n=4$. If one solves eqn. $\left(  \ref{eq:x&y-AC}\right)  $ for $y<y_{0}$ in
terms of $x$, there are always two solutions $x_{S}$ and $x_{L}$, where
$x_{S}<x_{0}<x_{L}.$ However, only the solution $x_{S}$ is Taylor expandable
in $\eta$. The solution $x_{L}$ is a "runaways" solution since it does not
exist in the limit of $\eta\rightarrow0$. In \cite{TBHRG-Simon:1990ic}, it has
been argued that "runaways" solutions are not physical and hence should be
discarded. A similar argument was also given in the framework of the
Generalized Uncertainty Principle in \cite{TBHRG-Ching:2012fu}. Therefore, we
will discard "runaways" solutions and keep only the solution $x_{S}$ in this
paper. In this case, we have $x_{cr}=x_{0}$ instead of $x_{cr}=\infty$. We
list $x_{cr}$ and $y_{cr}$ for various choices of $n$ and $\eta$ in TABLE
\ref{tab:one}.

To calculate the black hole's temperature from eqn. $\left(  \ref{eq:BH-Temp}%
\right)  \,$, we need to solve eqn. $\left(  \ref{eq:x&y-AC}\right)  $ for
$h\left(  y\right)  $. Nevertheless, it is non-trivial to get $h\left(
y\right)  $ for a general $n$. For $n=1,2,4$, we have
\begin{equation}
h\left(  y\right)  =\left\{
\begin{array}
[c]{l}%
\frac{1}{2}\left(  -\eta y+\sqrt{4+\eta^{2}y^{2}}\right)  \text{
\ \ \ \ }n=1\\
\text{ \ \ \ \ \ }\left(  1+\eta y^{2}\right)  ^{-1/2}\text{ \ \ \ \ }%
\ \ \ \ \ \ n=2\\
\text{ \ \ \ }y^{-2}\sqrt{\frac{\sqrt{1+4\eta y^{4}}-1}{2\eta}}\text{
\ \ \ \ \ \ \ \ \ }n=3
\end{array}
\right.  .
\end{equation}
If $y\ll1$, one has $x\ll1$ and hence eqn. $\left(  \ref{eq:x&y-AC}\right)  $
becomes%
\begin{equation}
y=x\left(  1+\frac{\eta x^{n}}{2}+\mathcal{O}\left(  x^{2n}\right)  \right)  ,
\end{equation}
which gives%
\begin{equation}
h\left(  y\right)  =1-\frac{\eta y^{n}}{2}+\mathcal{O}\left(  y^{2n}\right)  .
\label{eq:hFunction-Small}%
\end{equation}

Since $x=E/m_{p}$ and $y=p/m_{p}$, the AC dispersion relation for a massless
particle $\left(  \ref{eq:x&y-AC}\right)  $ leads to the group velocity of the
particle%
\begin{equation}
\frac{1}{v_{g}}=\frac{\partial y}{\partial x}=\frac{1}{\sqrt{1-\eta x^{n}}%
}+\frac{n\eta x^{n}}{2\left(  1-\eta x^{n}\right)  ^{3/2}}.
\label{eq:Group-Velocity}%
\end{equation}
If $\eta>0$, eqn. $\left(  \ref{eq:Group-Velocity}\right)  $ gives $v_{g}<1$,
which means the particle is subluminal. If $\eta<0$, $v_{g}>1$ and hence the
particle is superluminal.

We now calculate the temperatures and entropies of a Schwarzschild black hole
and a static uncharged black string with negative cosmological constant.

\bigskip

\begin{table}[pbh]
\begin{center}%
\begin{tabular}
[c]{|c|c|c|c|c|c|c|c|}\hline
& $x_{cr}$ & $y_{cr}$ & $M_{cr}^{SC}$ & $T_{cr}^{SC}/m_{p}$ & $M_{cr}^{BS}$ &
$T_{cr}^{BS}/m_{p}$ & Lines in figures\\\hline
$\eta=0$ & $\infty$ & $\infty$ & $0$ & $\infty$ & $0$ & $0$ & Blue
Solid\\\hline
$\eta>0$ & $\eta^{-1/n}$ & $\infty$ & $0$ & $\frac{\eta^{-1/n}}{4\pi}$ & $0$ &
$0$ & Black Solid\\\hline
$\eta<0,0<n<1$ & $\infty$ & $\infty$ & $0$ & $\infty$ & $0$ & $\infty$ &
\\\hline
$\eta<0,n=1$ & $\infty$ & $\infty$ & $0$ & $\infty$ & $0$ & $-\eta
\frac{3\alpha^{2}m_{p}^{2}}{4\pi}$ & Black Dashed\\\hline
$\eta<0,1<n<2$ & $\infty$ & $\infty$ & $0$ & $\infty$ & $0$ & $0$ & \\\hline
$\eta<0,n=2$ & $\infty$ & $\left(  -\eta\right)  ^{-1/2}$ & $\frac{m_{p}%
}{2\left(  -\eta\right)  ^{-1/2}}$ & $\infty$ & $\frac{\alpha^{3}m_{p}^{3}%
}{4\left(  -\eta\right)  ^{-3/2}}$ & $\infty$ & Red Dashed\\\hline
$\eta<0,n>2$ & $x_{0}$ & $y_{0}$ & $\frac{m_{p}}{2y_{0}}$ & $\frac{x_{0}}%
{4\pi}$ & $\frac{\alpha^{3}m_{p}^{3}}{4y_{0}^{3}}$ & $\frac{3\alpha^{2}%
m_{p}^{2}}{4\pi}\frac{x_{0}}{y_{0}^{2}}$ & Red Solid\\\hline
\end{tabular}
\end{center}
\caption{The values of $x_{cr}$, $y_{cr}$, $M_{cr}^{SC}$, $T_{cr}^{SC}/m_{p}$,
$M_{cr}^{BS}$, and $T_{cr}^{BS}/m_{p}$ for various values of $\left(
n,\eta\right)  $. The superscripts $SC$ and $BS$ stand for the Schwarzschild
black hole and black string, respectively.}%
\label{tab:one}%
\end{table}

\subsection{Schwarzschild Black Hole}

\begin{figure}[tb]
\begin{centering}
\includegraphics[scale=1.2]{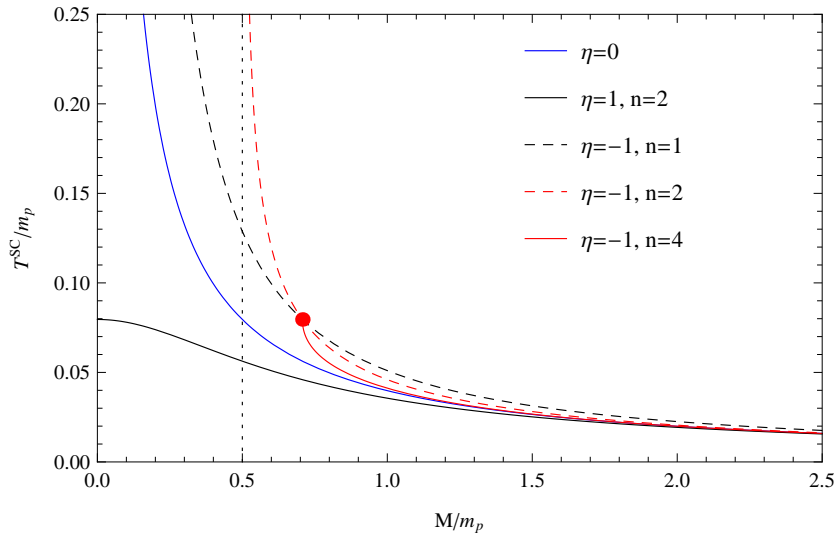}
\par\end{centering}
\caption{Plot of the temperature $T^{SC}/m_{p}$ against the mass $M/m_{p}$ for
a Schwarzschild black hole. All the lines asymptotically approach $T^{SC}=0$
as $M/m_{p}\rightarrow\infty$. The blue line is usual case, where $T^{SC}$
blows up as $M\rightarrow0$. The red dot is the end of the red solid line,
where the black hole has a remnant $M_{cr}^{SC}=1/\sqrt{2}m_{p}$. In this
case, $T^{SC}$ does not blow up as $M\rightarrow M_{cr}^{SC}$. The black
dotted line is the asymptotic line of the red dashed line as $M\rightarrow
M_{cr}^{SC}=0.5m_{p}$, which is the black hole's remnant. In this case,
$T^{SC}$ blows up as $M\rightarrow M_{cr}^{SC}$.}%
\label{fig:TBH}%
\end{figure}

For a Schwarzschild black hole with mass $M$, one has $B\left(  r\right)
=1-\frac{2M}{r}$ and $r_{h}=2M$. Thus, the temperature is%
\begin{equation}
T^{SC}=\frac{m_{p}^{2}}{8\pi M}h\left(  \frac{m_{p}}{2M}\right)
.\label{eq:Temp-SC}%
\end{equation}
If $M\gg m_{p},$ we have from eqn. $\left(  \ref{eq:hFunction-Small}\right)  $
that%
\begin{equation}
T^{SC}=\frac{m_{p}^{2}}{8\pi M}\left[  1-\frac{\eta}{2^{n+1}}\frac{m_{p}^{n}%
}{M^{n}}+\mathcal{O}\left(  \frac{m_{p}^{2n}}{M^{2n}}\right)  \right]  .
\end{equation}
The minimum mass $M_{cr}^{SC}$ of the black hole is given by%
\begin{equation}
M_{cr}^{SC}=\frac{m_{p}}{2y_{cr}}.\label{eq:T-BH}%
\end{equation}
When the mass $M$ reaches $M_{cr}^{SC},$ the temperature of the black hole is
denoted by $T_{cr}^{SC}$. Eqn. $\left(  \ref{eq:Tem-SCBH}\right)  $ gives that%
\begin{equation}
T_{cr}^{SC}=\frac{x_{cr}m_{p}}{4\pi}.\label{eq:Tem-SCBH}%
\end{equation}
For $\eta<0$ and $n\geq2$, $y_{cr}$ is finite and hence the black hole should
have non-vanishing minimum mass $M_{cr}^{SC}$. This implies an existence of
black hole remnant due to rainbow gravity. By eqn. $\left(  \ref{eq:Tem-SCBH}%
\right)  $, we find that $T_{cr}^{SC}$ for $n=2$ is infinite while
$T_{cr}^{SC}$ for $n>2$ is $\frac{x_{0}m_{p}}{4\pi}$, which is finite. For
$\eta<0$ and $0<n<2$, we find that $M_{cr}^{SC}=0$ and $T_{cr}^{SC}=\infty$.
In this case, the black hole would evaporate completely while its temperature
increases and finally becomes infinity during evaporation, just like the
standard Hawking radiation. For $\eta>0,$ the black hole would also evaporate
completely. However, the temperature of the black hole is a finite value
$\frac{\eta^{-1/n}m_{p}}{4\pi}$ at the end of the evaporation process. We list
$M_{cr}^{SC}$ and $T_{cr}^{SC}$ for all the possible values of $\eta$ and $n$
in TABLE \ref{tab:one}. In FIG. $\ref{fig:TBH}$, we plot the temperature
$T^{SC}/m_{p}$ against the black hole mass $M/m_{p}$(both in Planck units),
for examples with $\left(  \eta,n\right)  =\left(  1,2\right)  $, $\left(
\eta,n\right)  =\left(  -1,1\right)  $, $\left(  \eta,n\right)  =\left(
-1,2\right)  $, and $\left(  \eta,n\right)  =\left(  -1,4\right)  $. The
standard Hawking radiation is also plotted as a blue line in FIG.
$\ref{fig:TBH}$.

\begin{figure}[tb]
\centering
\begin{minipage}[b]{0.45\linewidth}
\includegraphics[width=\linewidth]{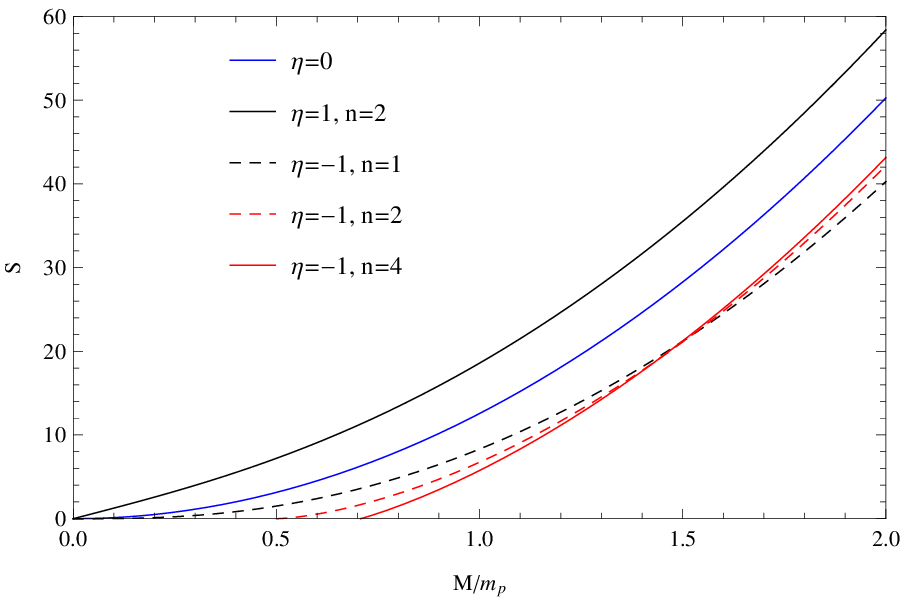}
\caption{\label{fig:SBH} Plot of the entropy $S$ against the mass $M/m_{p}$ for a Schwarzschild black hole. \\：}
\end{minipage}
\quad\begin{minipage}[b]{0.45\linewidth}
\includegraphics[width=\linewidth]{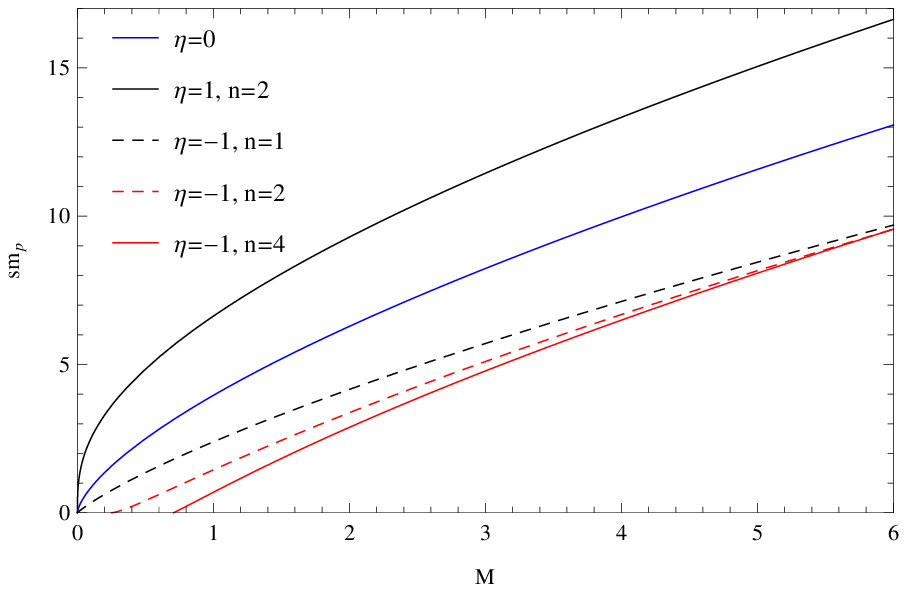}
\caption{\label{fig:SBS} Plot of the entropy of a Plank length $sm_{p}$ against the mass per unit length $M$ for a static uncharged black string.}
\end{minipage}
\end{figure}

From the first law of black hole thermodynamics $dS=dM/T^{SC}$, the entropy
associated with the temperature $\left(  \ref{eq:Temp-SC}\right)  $ can be
given by%
\begin{equation}
S=\int_{M_{cr}^{SC}}^{M}\frac{dM}{T^{SC}}=2\pi\int_{\frac{m_{p}}{2M}}^{y_{cr}%
}\frac{dy}{y^{3}h\left(  y\right)  }, \label{eq:Entropy-SC}%
\end{equation}
where we use $y_{cr}=\frac{m_{p}}{2M_{cr}^{SC}}$. For $\eta=0,$ we have
$h\left(  y\right)  =1$ and $y_{cr}=\infty$. Thus, eqn. $\left(
\ref{eq:Entropy-SC}\right)  $ gives the Bekenstein-Hawking entropy%
\begin{equation}
S=\frac{4\pi M^{2}}{m_{p}^{2}}=\frac{A}{4\hbar}.
\end{equation}
where $A=4\pi r_{h}^{2}$ is the area of the black hole. If $M\gg m_{p}\left(
A\gg\hbar\right)  $, eqn. $\left(  \ref{eq:Entropy-SC}\right)  $ gives the
entropy up to the subleading term
\begin{equation}
S\sim\left\{
\begin{array}
[c]{l}%
\frac{A}{4\hbar}+\frac{\pi\eta}{2-n}\left(  \frac{A}{4\pi\hbar}\right)
^{\frac{2-n}{2}}\text{ \ }0<n<2\\
\text{ \ }\frac{A}{4\hbar}+\frac{\pi\eta}{2}\ln\frac{A}{4\pi\hbar}\text{
\ \ \ \ \ \ \ \ \ \ }n=2\\
\text{ \ \ \ \ }\frac{A}{4\hbar}+S_{0}\text{ \ \ \ \ \ \ \ \ \ \ \ \ \ \ \ }%
n>2
\end{array}
\right.  , \label{eq:Entropy-SC-Small}%
\end{equation}
where $S_{0}$ is a constant and we use eqn. $\left(  \ref{eq:hFunction-Small}%
\right)  $. The leading terms of eqn. $\left(  \ref{eq:Entropy-SC-Small}%
\right)  $ are the familiar Bekenstein-Hawking entropy. For $n=2$, we obtain
the logarithmic subleading term. In FIG. $\ref{fig:SBH}$, we plot the entropy
$S$ against the black hole mass $M/m_{p}$, for examples with $\eta=0$,
$\left(  \eta,n\right)  =\left(  1,2\right)  $, $\left(  \eta,n\right)
=\left(  -1,1\right)  $, $\left(  \eta,n\right)  =\left(  -1,2\right)  $, and
$\left(  \eta,n\right)  =\left(  -1,4\right)  $.

In \cite{IN-Gim:2014ira}, the temperature and entropy of a rainbow
Schwarzschild black hole has also been computed in the case with $\eta>0$ and
$n=2$. The temperature and logarithmic subleading term of the entropy obtained
in our paper are the same as those in \cite{IN-Gim:2014ira}. By contrast, the
temperature and entropy were calculated in the case with $\eta>0$ in
\cite{IN-Ali:2014xqa}, where it was found that the black hole had a
non-vanishing minimum mass and the temperature was zero at the minimum mass.
As pointed out in \cite{IN-Gim:2014ira}, the difference between the results in
our paper and \cite{IN-Gim:2014ira} and those in \cite{IN-Ali:2014xqa} stems
from the fact the MDR was partially used in \cite{IN-Ali:2014xqa}. In fact,
the author in \cite{IN-Ali:2014xqa} used the MDR to get $T_{eff}$ while the
ordinary dispersion relation rather than the MDR was used to obtain the
relation between the energy $E$ of an emitted particle and the event horizon
radius $r_{h}$.

\subsection{Static Uncharged Black String}

We consider the Einstein-Hilbert action in four dimensions with a negative
cosmological constant $\Lambda$, which is given by%
\begin{equation}
S=\frac{1}{16\pi}\int d^{4}x\sqrt{-g}\left(  R-2\Lambda\right)  ,
\end{equation}
where $g$ is the determinant of the metric and $R$ the Ricci scalar. If we
assume that the solution of the Einstein equations is cylindrically symmetric
and time-independent, the line element for a static uncharged black string
becomes\cite{TBHRG-Lemos:1994xp}%
\begin{equation}
ds^{2}=\left(  \alpha^{2}r^{2}-\frac{b}{\alpha r}\right)  dt^{2}-\left(
\alpha^{2}r^{2}-\frac{b}{\alpha r}\right)  ^{-1}dr^{2}-r^{2}\left(
d\theta^{2}+\alpha^{2}dz^{2}\right)  , \label{eq:BSMetric}%
\end{equation}
where%
\begin{gather}
\alpha^{2}=-\frac{\Lambda}{3},\text{ }b=4M\nonumber\\
-\infty<t<\infty,\text{ }0\leq r<\infty,\text{ }-\infty<z<\infty,\text{ }%
0\leq\theta\leq2\pi.
\end{gather}
Here, $M$ is the mass per unit length of the $z$ line of black string.

\begin{figure}[tb]
\begin{centering}
\includegraphics[scale=1.2]{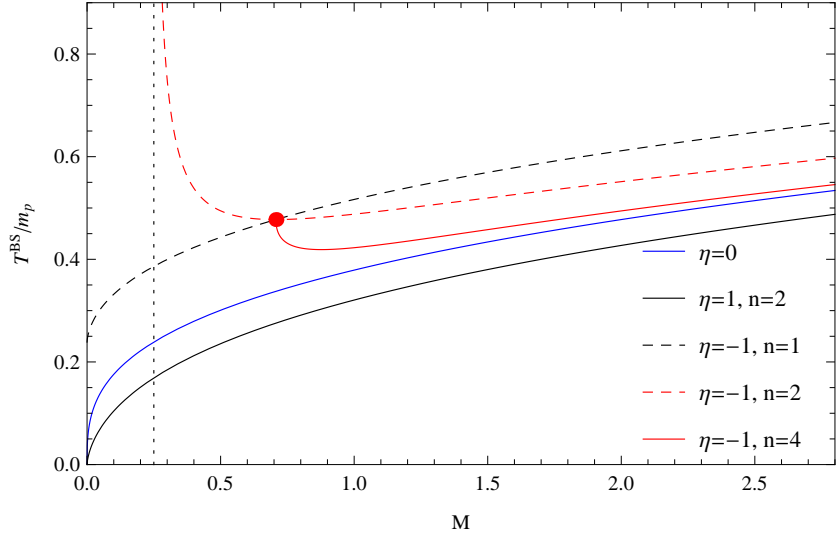}
\par\end{centering}
\caption{ Plot of the temperature $T^{BS}/m_{p}$ against the mass per unit
length $M$ for a static uncharged black string. The lines with $n>1$
asymptotically approach the blue line as $M\rightarrow\infty$. The blue line
is the usual case, where $T^{SC}\rightarrow0$ as $M\rightarrow0$. The red dot
is the end of the red solid line, where the black string has a remnant
$M_{cr}^{BS}=1/\sqrt{2}$. The black dotted line is the asymptotic line of the
red dashed line as $M\rightarrow M_{cr}^{BS}=0.25$, which is the black
string's remnant. In this case, $T^{BS}$ blows up as $M\rightarrow M_{cr}%
^{BS}$.}%
\label{fig:TBS}%
\end{figure}

For the static uncharged black string, we have $B\left(  r\right)  =\alpha
^{2}r^{2}-\frac{b}{\alpha r}$, $r_{h}=b^{1/3}/\alpha$, and $\kappa
=\frac{3\alpha}{2}b^{1/3}$. Thus, eqn. $\left(  \ref{eq:BH-Temp}\right)  $
leads to the black string temperature%
\begin{equation}
\frac{T^{BS}}{m_{p}}=\frac{3\alpha m_{p}b^{1/3}}{4\pi}h\left(  \frac{\alpha
m_{p}}{b^{1/3}}\right)  . \label{eq:Temp-BSh}%
\end{equation}
If $b^{1/3}\gg\alpha m_{p}\left(  M\gg\alpha^{3}m_{p}^{3}\right)  ,$ we have
for the temperature%
\begin{equation}
T^{BS}=\frac{3\alpha m_{p}b^{1/3}}{4\pi}\left[  1-\frac{\eta}{2^{n+1}}%
\frac{\alpha^{n}m_{p}^{n}}{b^{n/3}}+\mathcal{O}\left(  \frac{\alpha^{2n}%
m_{p}^{2n}}{b^{2n/3}}\right)  \right]  .
\end{equation}
If $n>1,\,$one has
\[
T^{BS}\rightarrow T_{0}\text{ \ as \ }M\rightarrow\infty,
\]
where $T_{0}=\frac{3\alpha m_{p}b^{1/3}}{4\pi}$ is the black string
temperature in the usual case. Since $r_{h}\geq\frac{m_{p}}{y_{cr}}$, the
minimum mass per unit length $M_{cr}^{BS}$ of the black string is given by%
\begin{equation}
M_{cr}^{BS}=\frac{\alpha^{3}m_{p}^{3}}{4y_{cr}^{3}}.
\end{equation}
When $M$ reaches $M_{cr}^{SC},$ the temperature of the black string becomes%
\begin{equation}
\frac{T_{cr}^{BS}}{m_{p}}=\frac{3\alpha^{2}m_{p}^{2}}{4\pi}\frac{x_{cr}%
}{y_{cr}^{2}}. \label{eq:Temp-BS}%
\end{equation}
For the usual case with $\eta=0$, we have $h\left(  y\right)  =1$ and
$y_{cr}=\infty$. The minimum mass per unit length $M_{cr}^{BS}=0$ and $T^{BS}$
goes to zero as $M\rightarrow0$. For $\eta<0$ and $n\geq2$, $y_{cr}$ is finite
and hence the black string should have non-vanishing minimum mass per unit
length $M_{cr}^{BS}$, which indicates an existence of black string remnant.
Using eqn. $\left(  \ref{eq:Temp-BS}\right)  $, we can show that $T_{cr}^{BS}$
for $n=2$ is infinity while $T_{cr}^{BS}$ for $n>2$ is finite. For $\eta>0,$
the black string would evaporate completely with $T^{BS}$ goes to zero as
$M\rightarrow0$, which is the same as in the usual case. For $\eta<0$ and
$0<n<2$, the minimum mass per unit length $M_{cr}^{BS}$ is zero. Nevertheless,
we find that $T_{cr}^{BS}=\infty$ for $0<n<1$, $T_{cr}^{BS}=$ $-\eta
\frac{3\alpha^{2}m_{p}^{3}}{4\pi}$ for $n=1$, and $T_{cr}^{BS}=0$ for $0$. We
list $M_{cr}^{BS}$ and $T_{cr}^{BS}$ for various values of $\eta$ and $n$ in
TABLE \ref{tab:one}. In FIG. $\ref{fig:TBS}$ where we set $\alpha m_{p}=1$, we
plot the temperature $T^{BS}/m_{p}$ against the mass per unit length $M$, for
examples with $\eta=0$, $\left(  \eta,n\right)  =\left(  1,2\right)  $,
$\left(  \eta,n\right)  =\left(  -1,1\right)  $, $\left(  \eta,n\right)
=\left(  -1,2\right)  $, and $\left(  \eta,n\right)  =\left(  -1,4\right)  $.

Using the first law of black hole thermodynamics, we have for the entropy per
unit length associated with the temperature $\left(  \ref{eq:Temp-BSh}\right)
$%
\begin{equation}
s=\int_{M_{cr}^{SC}}^{M}\frac{dM}{T^{SC}}=\frac{\pi\alpha m_{p}}{m_{p}}%
\int_{\frac{\alpha m_{p}}{\left(  4M\right)  ^{1/3}}}^{y_{cr}}\frac{dy}%
{y^{3}h\left(  y\right)  }. \label{eq:Entropy-BS}%
\end{equation}
If $M\gg1$, eqn. $\left(  \ref{eq:Entropy-BS}\right)  $ gives the entropy per
unit length up to the subleading term%
\begin{equation}
s\sim\left\{
\begin{array}
[c]{l}%
\frac{\pi\alpha}{2}\frac{b^{2/3}}{\alpha^{2}m_{p}^{2}}+\frac{\alpha\pi\eta
}{2\left(  2-n\right)  }\frac{b^{\left(  2-n\right)  /3}}{\left(  \alpha
m_{p}\right)  ^{2-n}}\text{ \ }0<n<2\\
\text{ \ }\frac{\pi\alpha}{2}\frac{b^{2/3}}{\alpha^{2}m_{p}^{2}}+\frac
{\pi\alpha\eta}{2}\ln\frac{b^{1/3}}{\alpha m_{p}}\text{ \ \ \ \ \ \ \ }n=2\\
\text{ \ \ \ \ \ }\frac{\pi\alpha}{2}\frac{b^{2/3}}{\alpha^{2}m_{p}^{2}}%
+s_{0}\text{ \ \ \ \ \ \ \ \ \ \ \ \ \ \ }n>2
\end{array}
\right.  ,
\end{equation}
where $s_{0}$ is a constant. In FIG. $\ref{fig:SBH}$, we plot $m_{p}S$ against
$M$, for examples with $\eta=0$, $\left(  \eta,n\right)  =\left(  1,2\right)
$, $\left(  \eta,n\right)  =\left(  -1,1\right)  $, $\left(  \eta,n\right)
=\left(  -1,2\right)  $, and $\left(  \eta,n\right)  =\left(  -1,4\right)  $.

\section{Luminosities of Black Holes in Rainbow Gravity}

\label{Sec:LBHRG}

For particles emitted in a wave mode labelled by energy $E$ and quantum
numbers $i$, we find that%
\begin{align*}
&  \left(  \text{Probability for a black hole to emit a particle in this
mode}\right) \\
&  =\exp\left(  -\frac{E}{T_{eff}}\right)  \times(\text{Probability for a
black hole to absorb a particle in the same mode}),
\end{align*}
where $T_{eff}$ is given by eqn. $\left(  \ref{eq:Eff-Temp}\right)  $. The
above relation for the usual dispersion relation was obtained by Hartle and
Hawking \cite{LBHRG-Hartle:1976tp} using semiclassical analysis. Neglecting
back-reaction, detailed balance condition requires that the ratio of the
probability of having $N$ particles in a particular mode to the probability of
having $N-1$ particles in the same mode is $\exp\left(  -\frac{E}{T_{eff}%
}\right)  .$ One then follows the argument in \cite{SF} to get the average
number $n_{E,i}$ in the mode with $E$ and $i$%
\begin{equation}
n_{E,i}=n\left(  \frac{E}{T_{eff}}\right)  , \label{eq:Number}%
\end{equation}
where we define%
\begin{equation}
n\left(  x\right)  =\frac{1}{\exp x-\left(  -1\right)  ^{\epsilon}}.
\end{equation}
Note that $\epsilon=0$ for bosons and $\epsilon=1$ for fermions.

As discussed in section \ref{Sec:TBHRG}, there is an upper bound $m_{p}x_{cr}$
on the energy $E$ of the particle. Moreover, another upper bound comes from
the requirement that nothing can be emitted that lowers the energy below the
remnant mass of a black hole. Thus, one has $E\leq M-M_{cr},$ where $M$ is the
mass of the black hole and $M_{cr}$ is the remnant mass. Considering both
upper bounds, we have for $E$%
\begin{equation}
E\leq E_{\max}\equiv\min\left\{  m_{p}x_{cr},M-M_{cr}\right\}  .
\label{eq:Emax}%
\end{equation}

In \cite{SF,LBHRG-Mu:2015qta,FF}, we have found that the total luminosity in
the MDR case is given by
\begin{equation}
L=%
{\displaystyle\sum\limits_{i}}
\int\left\vert T_{i}\left(  E\right)  \right\vert ^{2}\omega n_{E,i}\frac
{dE}{2\pi\hbar}\text{\thinspace}. \label{eq:Luminocity}%
\end{equation}
where $E$ is the energy of the particle, $i$ are quantum numbers needed to
specify a mode besides $E$, and $\left\vert T_{i}\left(  E\right)  \right\vert
^{2}$ is the greybody factor. Usually, $T_{i}\left(  E\right)  $ represents
the transmission coefficient of the black hole barrier which in general can
depend on $E$ and $i$. The relevant radiations usually have the energy of
order $\hbar M^{-1}$ for a black hole with the mass $M$ and one hence needs to
use the wave equations to compute $T_{i}\left(  E\right)  $. However, solving
the wave equations for $T_{i}\left(  E\right)  $ could be complex. On the
other hand, we can use the geometric optics approximation to estimate
$\left\vert T_{i}\left(  E\right)  \right\vert ^{2}$. In the geometric optics
approximation, we assume $E\gg M$ and high energy waves will be absorbed
unless they are aimed away from the black hole. Hence $\left\vert T_{i}\left(
E\right)  \right\vert ^{2}=1$ for all the classically allowed $E$ and $i$,
while $\left\vert T_{i}\left(  E\right)  \right\vert ^{2}=0$ otherwise. For
the usual dispersion relation, the well-known Stefan's law for black holes is
obtained in this approximation.

In the remaining of the section, we will use the geometric optics
approximation to calculate luminosities of a 2D rainbow black hole, a 4D
rainbow spherically symmetric one, and a 4D rainbow cylindrically symmetric
one. For simplicity, we assume that the particles are massless.

\subsection{2D Black Hole}

Suppose the metric of a 2D black hole is given by%
\begin{equation}
ds^{2}=B\left(  r\right)  dt^{2}-\frac{1}{B\left(  r\right)  }dr^{2},
\end{equation}
where $B\left(  r\right)  $ has a simple zero at $r=r_{h}$. In this case, we
have $\left\vert T\left(  E\right)  \right\vert ^{2}=1$ when $E\leq E_{\max}$
and $\left\vert T\left(  E\right)  \right\vert ^{2}=0$ otherwise. By eqn.
$\left(  \ref{eq:Luminocity}\right)  $, the luminosity of a 2D black hole is%
\begin{equation}
L=g_{s}\int_{0}^{E_{\max}}En\left(  \frac{E}{T_{0}}\frac{f\left(
E/m_{p}\right)  }{g\left(  E/m_{p}\right)  }\right)  \frac{dE}{2\pi\hbar
},\label{eq:L-2D}%
\end{equation}
where we use eqn. $\left(  \ref{eq:Eff-Temp}\right)  $ for $T_{eff}$ and
$T_{0}=\frac{m_{p}^{2}\kappa}{2\pi}$ with $\kappa=\frac{B^{\prime}\left(
r_{h}\right)  }{2}$. Note $g_{s}$ is the number of polarization which is $1$
for scalars and $2$ for spin-$1/2$ fermion and vector bosons. Defining
\begin{equation}
u=\frac{E}{T_{0}}\frac{f\left(  E/m_{p}\right)  }{g\left(  E/m_{p}\right)
},\label{eq:uandE}%
\end{equation}
we find%
\begin{equation}
E=T_{0}uh\left(  \frac{uT_{0}}{m_{p}}\right)  .
\end{equation}
Using eqn. $\left(  \ref{eq:uandE}\right)  $ to change variables in eqn.
$\left(  \ref{eq:L-2D}\right)  $, we have for the luminosity%
\begin{equation}
L=\frac{g_{s}T_{0}^{2}}{2\pi m_{p}^{2}}\int_{0}^{u_{\max}}h\left(
\frac{uT_{0}}{m_{p}}\right)  \left[  h\left(  \frac{uT_{0}}{m_{p}}\right)
+\left(  \frac{uT_{0}}{m_{p}}\right)  h^{\prime}\left(  \frac{uT_{0}}{m_{p}%
}\right)  \right]  un\left(  u\right)  du,\label{eq:L-2d-u}%
\end{equation}
where we define%
\begin{equation}
u_{\max}=\frac{E_{\max}}{T_{0}}\frac{f\left(  \frac{E_{\max}}{m_{p}}\right)
}{g\left(  \frac{E_{\max}}{m_{p}}\right)  }.
\end{equation}
Note that the change of variables given in eqn. $\left(  \ref{eq:uandE}%
\right)  $ is legitimate for the integral $\left(  \ref{eq:L-2d-u}\right)  $
if the function $\frac{xf\left(  x\right)  }{g\left(  x\right)  }$ is
monotonic over $\left(  0,x_{cr}\right)  $, which is the case for the AC
dispersion relation.

Now investigate properties of the luminosity for the AC dispersion relation
given in eqn. $\left(  \ref{eq:AC-Dispersion}\right)  $. If $T_{0}\ll m_{p}$,
eqn. $\left(  \ref{eq:L-2d-u}\right)  $ becomes
\begin{equation}
L\approx\frac{g_{s}T_{0}^{2}}{2\pi m_{p}^{2}}\int_{0}^{\infty}\left[
1-\frac{n+2}{2}\eta\left(  \frac{uT_{0}}{m_{p}}\right)  ^{n}\right]  un\left(
u\right)  du
\end{equation}
where we set $u_{\max}=\infty$. For the emission of $n_{s}$ species of
massless particles of spin $s,$ we have%
\begin{align}
L &  \approx\frac{\pi}{12}\left(  \frac{T_{0}}{m_{p}}\right)  ^{2}\left\{
\left(  n_{0}+2n_{1}+n_{1/2}\right)  \right.  \nonumber\\
&  \left.  -\frac{3\left(  n+1\right)  !\left(  n+2\right)  }{\pi^{2}}\left(
\frac{T_{0}}{m_{p}}\right)  ^{n}\eta\left[  Li_{n+2}\left(  1\right)  \left(
n_{0}+2n_{1}\right)  +\frac{2-2^{-n}}{n+2}\zeta\left(  n+2\right)
n_{1/2}\right]  \right\}  ,
\end{align}
where $Li_{s}\left(  z\right)  $ is the polylogarithm of order $s$ and
argument $z$ and $\zeta\left(  z\right)  $ is the zeta function. If
$\eta>0\left(  \eta<0\right)  $, the luminosity becomes smaller(larger) than
that in the usual case with $\eta=0$. This is expected from eqn. $\left(
\ref{eq:Eff-Temp}\right)  $ where $T_{eff\text{ }}$is lowered(raised) due to
rainbow gravity if $\eta>0\left(  \eta<0\right)  $.

During the late stage of the black hole evaporation process when
$\frac{M-M_{cr}}{m_{p}}\ll\min\left\{  1,x_{cr}\right\}  $, one has
$\frac{E_{\max}}{m_{p}}=\frac{M-M_{cr}}{m_{p}}\ll1\,$\ and hence
\begin{equation}
u_{\max}\approx\frac{m_{p}}{T_{0}}\frac{M-M_{cr}}{m_{p}}\left[  1+\frac{\eta
}{2}\left(  \frac{M-M_{cr}}{m_{p}}\right)  ^{n}\right]  .
\end{equation}
The luminosity of radiation of one species of bosons is%
\begin{equation}
L=\frac{g_{s}T_{0}}{2\pi m_{p}}\frac{M-M_{cr}}{m_{p}}\left[  1-\frac{\eta}%
{2}\frac{1}{n+1}\left(  \frac{M-M_{cr}}{m_{p}}\right)  ^{n}+\mathcal{O}\left(
\left(  \frac{M-M_{cr}}{m_{p}}\right)  ^{2n}\right)  \right]
,\label{eq:L-2D-Small-Boson}%
\end{equation}
and that of radiation of one species of fermions is%
\begin{equation}
L=\frac{g_{s}}{4\pi}\left(  \frac{M-M_{cr}}{m_{p}}\right)  ^{2}\left[
1+\mathcal{O}\left(  \left(  \frac{M-M_{cr}}{m_{p}}\right)  ^{2n}\right)
\right]  ,\label{eq:L-2D-Small-Fermion}%
\end{equation}
where $g_{s}$ is the number of polarization. From eqns. $\left(
\ref{eq:L-2D-Small-Boson}\right)  $ and $\left(  \ref{eq:L-2D-Small-Fermion}%
\right)  $, we find that the black hole evaporates mostly via bosons in the
late stage of the black hole evaporation process.

\subsection{4D Spherically Symmetric Black Hole}

For a 4D spherically symmetric black hole with $h_{ab}\left(  x\right)
dx^{a}dx^{b}=d\theta^{2}+\sin^{2}\theta d\phi^{2}$ in eqn. $\left(
\ref{eq:BHmetric}\right)  $, we have found $\lambda$ in eqn. $\left(
\ref{eq:HJ-Lamda&W}\right)  $ is given by \cite{SF}%
\begin{equation}
\lambda=\left(  l+\frac{1}{2}\right)  ^{2}\hbar^{2},
\end{equation}
where $l=0,1,\cdots$ is the angular momentum. Since $p_{r}$ in eqn. $\left(
\ref{eq:HJ-Lamda&W}\right)  $ is always a real number in the geometric optics
approximation, one has an upper bound on $\lambda$%
\begin{equation}
\lambda\leq\frac{C\left(  r^{2}\right)  }{B\left(  r\right)  }\frac
{f^{2}\left(  E/m_{p}\right)  }{g^{2}\left(  E/m_{p}\right)  }E^{2}.
\end{equation}
Suppose $\frac{C\left(  r^{2}\right)  }{B\left(  r\right)  }$ has a minimum at
$r_{\min}$. If the particles overcome the angular momentum barrier and get
absorbed by the black hole, one has%
\begin{equation}
\lambda\leq\lambda_{\max}\equiv\frac{C\left(  r_{\min}^{2}\right)  }{B\left(
r_{\min}\right)  }\frac{f^{2}\left(  E/m_{p}\right)  }{g^{2}\left(
E/m_{p}\right)  }E^{2}.
\end{equation}
The luminosity is%
\begin{align}
L &  =g_{s}%
{\displaystyle\sum\limits_{l}}
\left(  2l+1\right)  \int En_{E,l}\frac{dE}{2\pi\hbar}\nonumber\\
&  =g_{s}\int_{0}^{E_{\max}}\frac{EdE}{2\pi\hbar^{3}}n\left(  \frac{E}{T_{0}%
}\frac{f\left(  E/m_{p}\right)  }{g\left(  E/m_{p}\right)  }\right)  \int
_{0}^{\lambda_{\max}}d\left[  \left(  l+\frac{1}{2}\right)  ^{2}\right]
\nonumber\\
&  =\frac{g_{s}}{2\pi\hbar^{3}}\frac{C\left(  r_{\min}^{2}\right)  }{B\left(
r_{\min}\right)  }\int_{0}^{E_{\max}}n\left(  \frac{E}{T_{0}}\frac{f\left(
E/m_{p}\right)  }{g\left(  E/m_{p}\right)  }\right)  \frac{f^{2}\left(
E/m_{p}\right)  }{g^{2}\left(  E/m_{p}\right)  }E^{3}dE,\label{eq:L-4D-SSBH}%
\end{align}
where we use eqn. $\left(  \ref{eq:Number}\right)  $ for $n_{E,l}$. Making
change of variables for eqn. $\left(  \ref{eq:L-4D-SSBH}\right)  $
\begin{equation}
u=\frac{E}{T_{0}}\frac{f\left(  E/m_{p}\right)  }{g\left(  E/m_{p}\right)  },
\end{equation}
the luminosity of a 4D spherically symmetric black hole becomes%
\begin{equation}
L=\frac{g_{s}T_{0}^{4}}{2\pi m_{p}^{4}}\frac{C\left(  r_{\min}^{2}\right)
}{B\left(  r_{\min}\right)  m_{p}^{2}}\int_{0}^{u_{\max}}h\left(  \frac
{uT_{0}}{m_{p}}\right)  \left[  h\left(  \frac{uT_{0}}{m_{p}}\right)  +\left(
\frac{uT_{0}}{m_{p}}\right)  h^{\prime}\left(  \frac{uT_{0}}{m_{p}}\right)
\right]  u^{3}n\left(  u\right)  du.\label{eq:L-4D-SSBH-u}%
\end{equation}

In what follows, we focus on the AC dispersion relation. If $T_{0}\ll m_{p}$,
eqn. $\left(  \ref{eq:L-4D-SSBH-u}\right)  $ becomes
\begin{equation}
L\approx\frac{g_{s}T_{0}^{4}}{2\pi m_{p}^{4}}\frac{C\left(  r_{\min}%
^{2}\right)  }{B\left(  r_{\min}\right)  m_{p}^{2}}\int_{0}^{\infty}\left[
1-\frac{n+2}{2}\eta\left(  \frac{uT_{0}}{m_{p}}\right)  ^{n}\right]
u^{3}n\left(  u\right)  du,
\end{equation}
where we set $u_{\max}=\infty$. For the emission of $n_{s}$ species of
massless particles of spin $s,$ we have%
\begin{align}
L &  \approx\frac{\pi^{3}}{30}\left(  \frac{T_{0}}{m_{p}}\right)  ^{4}%
\frac{C\left(  r_{\min}^{2}\right)  }{m_{p}^{2}B\left(  r_{\min}\right)
}\left\{  \left(  n_{0}+2n_{1}+\frac{7}{4}n_{1/2}\right)  \right.  \nonumber\\
&  \left.  -\frac{15\left(  n+3\right)  !\left(  n+2\right)  }{2\pi^{4}%
}\left(  \frac{T_{0}}{m_{p}}\right)  ^{n}\eta\left[  Li_{n+4}\left(  1\right)
\left(  n_{0}+2n_{1}\right)  +\left(  2-2^{-n-2}\right)  \zeta\left(
n+4\right)  n_{1/2}\right]  \right\}  ,\label{eq:L-4D-smallT}%
\end{align}
where $Li_{s}\left(  z\right)  $ is the polylogarithm of order $s$ and
argument $z$ and $\zeta\left(  z\right)  $ is the zeta function. In the late
stage of the black hole evaporation process with $\frac{M-M_{cr}}{m_{p}}%
\ll\min\left\{  1,x_{cr}\right\}  $, one finds that the luminosity of
radiation of one species of bosons is%
\begin{equation}
L=\frac{g_{s}T_{0}}{6\pi m_{p}}\frac{C\left(  r_{\min}^{2}\right)  }{m_{p}%
^{2}B\left(  r_{\min}\right)  }\left(  \frac{M-M_{cr}}{m_{p}}\right)
^{3}\left[  1+\frac{3\eta}{2\left(  n+3\right)  }\left(  \frac{M-M_{cr}}%
{m_{p}}\right)  ^{n}+\mathcal{O}\left(  \left(  \frac{M-M_{cr}}{m_{p}}\right)
^{2n}\right)  \right]  ,\label{eq:L-4D-LateStageB}%
\end{equation}
and that of radiation of one species of fermions%
\begin{equation}
L=\frac{g_{s}}{8\pi}\frac{C\left(  r_{\min}^{2}\right)  }{m_{p}^{2}B\left(
r_{\min}\right)  }\left(  \frac{M-M_{cr}}{m_{p}}\right)  ^{4}\left[
1+\frac{4\eta}{n+4}\left(  \frac{M-M_{cr}}{m_{p}}\right)  ^{n}+\mathcal{O}%
\left(  \left(  \frac{M-M_{cr}}{m_{p}}\right)  ^{2n}\right)  \right]
,\label{eq:L-4D-LateStageF}%
\end{equation}
where $g_{s}$ is the number of polarization. Similar to a 2D black hole, the
evaporation via bosons dominates the 4D spherically symmetric black hole
evaporation process in the late stage.

In the geometric optics approximation, a 4D spherically symmetric black hole
can be described as a black sphere for absorbing particles. The total
luminosity are determined by the radius of the black sphere $R$ and the
temperature of the black hole $T$. For the AC dispersion relation, we have for
massless particles
\begin{equation}
R=\sqrt{\frac{\lambda_{\max}}{E^{2}}}=\frac{C\left(  r_{\min}^{2}\right)
}{B\left(  r_{\min}\right)  }\frac{1}{1-\eta\left(  E/m_{p}\right)  ^{n}%
}\text{ and }T=T_{eff}=T_{0}\sqrt{1-\eta\left(  E/m_{p}\right)  ^{n}}.
\end{equation}
If $\eta>0\left(  \eta<0\right)  $, the radius $R$ becomes larger(smaller)
than that in the usual case while the effective temperature becomes
lower(higher) due to rainbow gravity. For the subluminal case with $\eta
>0$(the superluminal case with $\eta<0$), the competition between the
increased(decreased) radius and the decreased(increased) temperature
determines whether the luminosity would increase or decrease. For $T_{0}\ll
m_{p}$, it appears from eqn. $\left(  \ref{eq:L-4D-smallT}\right)  $ that the
effects of decreased(increased) temperature wins the competition and hence the
luminosity tends to become smaller(larger). However for the late stage of the
evaporation process, it seems from eqns. $\left(  \ref{eq:L-4D-LateStageB}%
\right)  $ and $\left(  \ref{eq:L-4D-LateStageF}\right)  $ that the effects of
increased(decreased) radius wins the competition and hence the luminosity
becomes larger(smaller).

\begin{figure}[tb]
\begin{centering}
\includegraphics[scale=1.2]{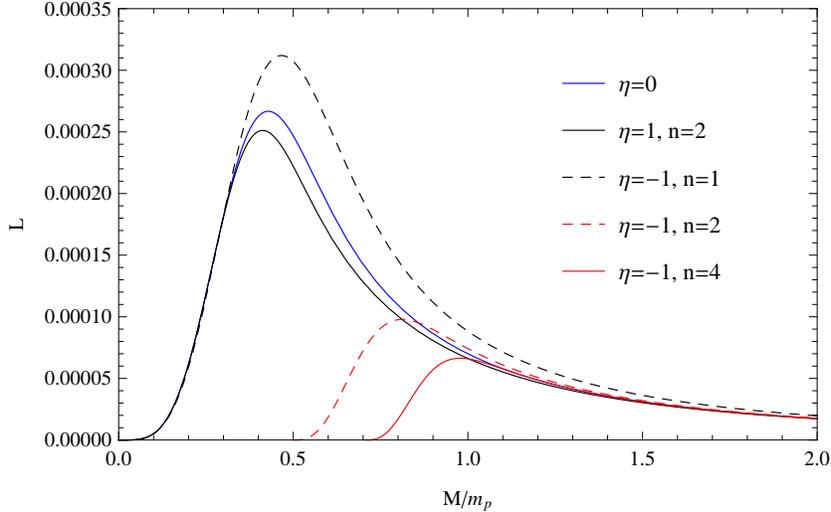}
\par\end{centering}
\caption{Plot of the luminosity $L$ of radiation of one species of bosons
against the mass $M/m_{p}$ for a Schwarzschild black hole. The blue line is
the usual case. All the other lines asymptotically approach the blue line as
$M\gg1$. For the cases with the minimal mass $M_{cr}^{SC}=0$ , the black lines
asymptotically approach the blue line as $M\rightarrow0$. For all the lines,
the luminosities approach zero as $M\rightarrow M_{cr}^{SC}$.}%
\label{fig:L4DBH}%
\end{figure}

We now work with a Schwarzschild black hole to investigate more properties of
the black hole's luminosity. For a Schwarzschild black hole, one has $B\left(
r\right)  =1-\frac{2M}{r},$ $r_{h}=2M,$ $\kappa=\frac{1}{4M}$, $T_{0}%
=\frac{m_{p}^{2}}{8\pi M}$, $r_{\min}=3M$ and $\frac{C\left(  r_{\min}%
^{2}\right)  }{B\left(  r_{\min}\right)  }=27M^{2}$. For $M\gg m_{p}$, it
shows form eqn. $\left(  \ref{eq:L-4D-smallT}\right)  $ that the corrections
to the luminosity from rainbow gravity effects are around $\mathcal{O}\left(
\frac{m_{p}}{M}\right)  ^{n}$. Nevertheless, these corrections begin to become
appreciable around $M\lesssim M_{\ast}\equiv\frac{c_{n}^{1/n}}{8\pi}\left\vert
\eta\right\vert ^{1/n}m_{p}$\ when the second term in the bracket of eqn.
$\left(  \ref{eq:L-4D-smallT}\right)  $ becomes comparable to 1. Here $c_{n}$
is the numerical factor in front of $n_{s}$ of the second term in the bracket
and $c_{n}^{1/n}\sim5-10$ for $1\leq n\leq10$. If $M_{cr}^{SC}=0$, in the late
stage of the evaporation process with $\frac{M}{m_{p}}\ll\min\left\{
1,x_{cr}\right\}  $, the corrections to the luminosity are around
$\mathcal{O}\left(  \frac{M^{n}}{m_{p}^{n}}\right)  $. Similarly, these
corrections are important when $M\gtrsim M_{\ast\ast}\equiv\left\vert
\eta\right\vert ^{-1/n}m_{p}$. Thus, we can conclude that for cases with
$M_{cr}^{SC}=0$, the effects of rainbow gravity impacts the black hole's
luminosity noticeably when $M_{\ast\ast}\lesssim M\lesssim M_{\ast}$. For
cases with non-vanishing $M_{cr}^{SC}$, the the black hole's luminosity
deviates from that in the usual case appreciably when $M\lesssim M_{\ast}$. In
FIG. \ref{fig:L4DBH}, we plot the luminosity $L$ of radiation of one species
of bosons against $M/m_{p}$ for examples with $\eta=0$, $\left(
\eta,n\right)  =\left(  1,2\right)  $, $\left(  \eta,n\right)  =\left(
-1,1\right)  $, $\left(  \eta,n\right)  =\left(  -1,2\right)  $, and $\left(
\eta,n\right)  =\left(  -1,4\right)  $. Note that the effects of rainbow
gravity does not change the black hole's luminosity appreciably in FIG.
\ref{fig:L4DBH} since $M_{\ast}\sim\left\vert \eta\right\vert ^{1/n}%
m_{p}=\left\vert \eta\right\vert ^{-1/n}m_{p}\sim M_{\ast\ast}$ for
$\left\vert \eta\right\vert =1$. Due to the requirement that the energy $E$ of
emitted particles could not exceed $M-M_{cr}^{SC}$, the luminosities in all
cases approach zero as $M\rightarrow M_{cr}^{SC}$.

\begin{figure}[tb]
\begin{center}
\subfigure[{~ Subluminal cases { } { } { }{ } { } { }{ } { } { }{ } { } { }{ } { } { }{ } { } { }{ } { } { }{ } { } { }{ } { } { }{ } { } { } $\left(\eta>0 \right)$}]{
\includegraphics[width=0.48\textwidth]{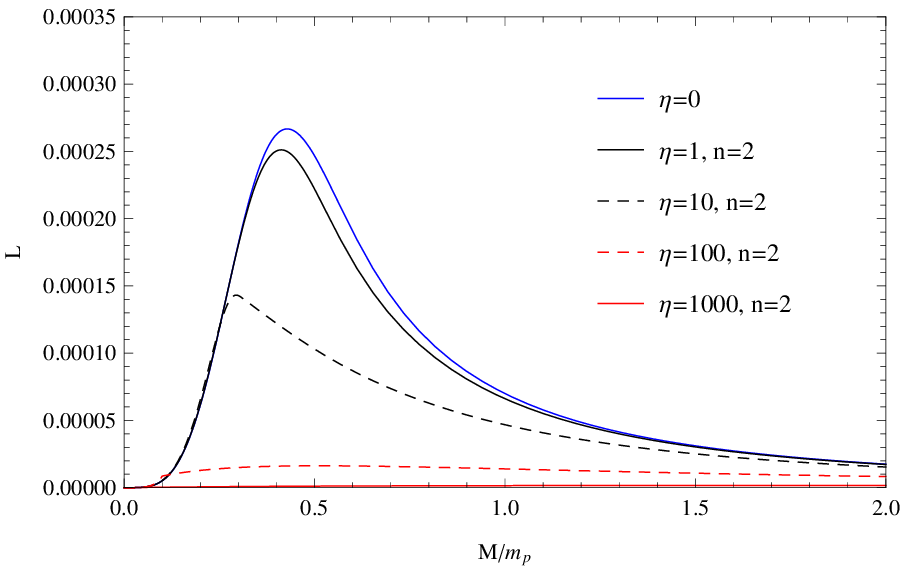}\label{fig:etaBH:a}}
\subfigure[{~Superluminal cases with remnant $\left(  \eta<0\text{ and }n\geq2\right)  $}]{
\includegraphics[width=0.48\textwidth]{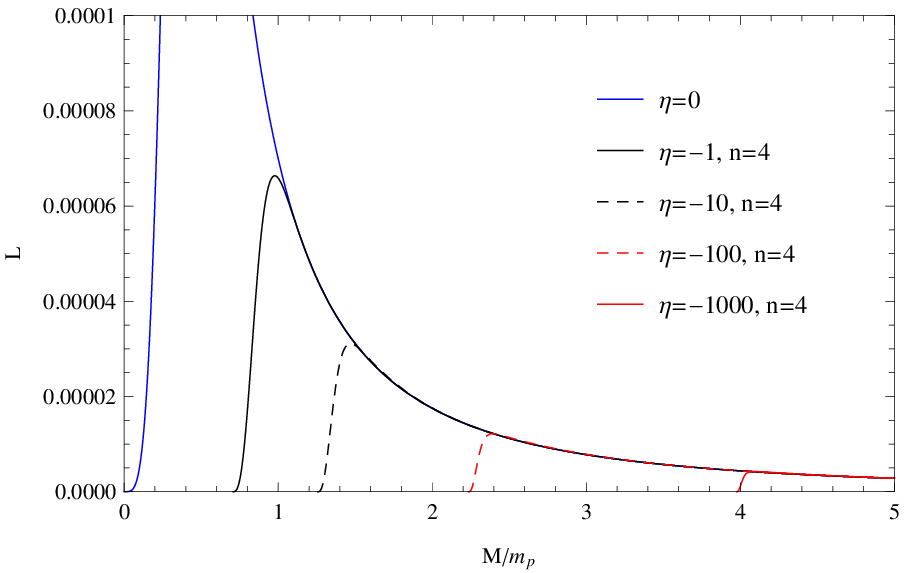}\label{fig:etaBH:b}}
\subfigure[{~Superluminal cases without remnant $\left(  \eta<0\text{ and }0<n<2\right)  $}]{
\includegraphics[width=0.48\textwidth]{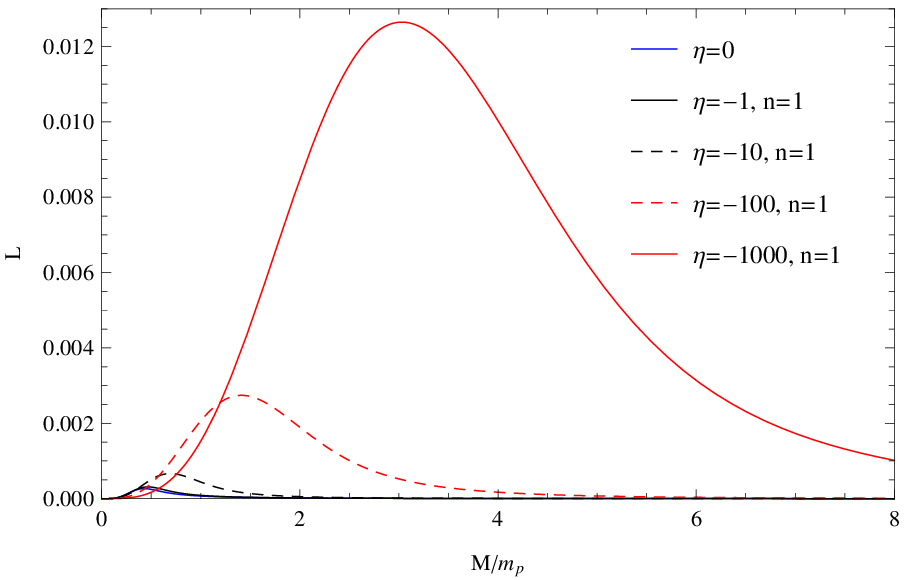}\label{fig:etaBH:c}}
\subfigure[{~Superluminal cases without remnant $\left(  \eta<0\text{ and }0<n<2\right)  $
}]{
\includegraphics[width=0.48\textwidth]{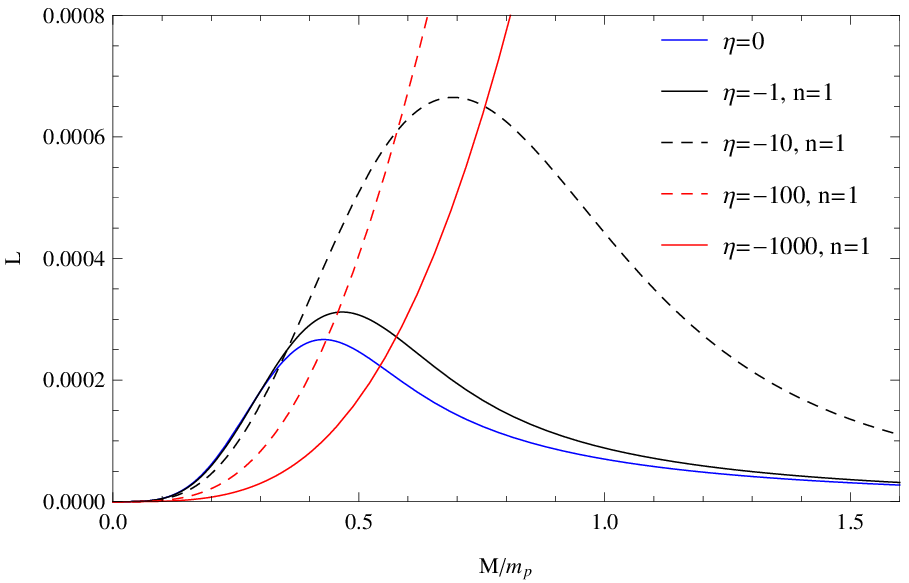}\label{fig:etaBH:d}}
\end{center}
\caption{Plots of the luminosity $L$ of radiation of one species of bosons
against the mass $M/m_{p}$ for a Schwarzschild black hole. The luminosity $L$
is plotted in subluminal cases and superluminal cases with and without remnant
for various values of $\eta$.}%
\label{fig:etaBH}%
\end{figure}

To study the sensitivity of the BH dynamics to the parameter $\eta$, we plot
the luminosity $L$ of radiation of one species of bosons against $M/m_{p}$ for
different values of $\eta$ in FIG. \ref{fig:etaBH}. Since $M_{\ast}%
\sim\left\vert \eta\right\vert ^{1/n}m_{p}$ and $M_{\ast\ast}\sim\left\vert
\eta\right\vert ^{-1/n}m_{p},$ we find that the larger $\left\vert
\eta\right\vert $ is, the more apparent the effects of rainbow gravity on the
luminosities become. In FIG. \ref{fig:etaBH}, we plot three cases as follows:

\begin{enumerate}
\item[$\left(  a\right)  $] Subluminal cases where $\eta>0$ and $M_{cr}%
^{SC}=0$. An example with $n=2$ is plotted in FIG. \ref{fig:etaBH:a} for
$\eta=1,10,100,$ and $1000$. The red solid line$\left(  \eta=1000\right)  $
can be barely seen since it is too close to the horizontal axis. It is evident
that the luminosity for $M_{\ast\ast}\lesssim M\lesssim M_{\ast}$ gets more
suppressed as the parameter $\eta$ becomes larger. In other words, the effects
of rainbow gravity could dramatically slow down the evaporation process of the
black hole for large enough $\eta$. Hence, the corresponding characteristic
time scale for the black hole to evaporate from $M_{\ast}$ to $M_{\ast\ast}$
could be much longer than in the usual case.

\item[$\left(  b\right)  $] Superluminal cases with the non-vanishing remnant
mass $M_{cr}^{SC}$ where $\eta<0$ and $n\geq2$. An example with $n=4$ is
plotted in FIG. \ref{fig:etaBH:b} for $\eta=-1,-10,-100,$ and $-1000$. In this
case, eqn. $\left(  \ref{eq:T-BH}\right)  $ gives that $M_{cr}^{SC}%
=\frac{m_{p}\left\vert \eta\right\vert ^{1/n}}{\sqrt{2}}\gtrsim M_{\ast}$. It
shows that the luminosity starts to deviate from that in the usual case when
$M$ is close to $M_{cr}^{SC}$ and then decreases to zero once $M=M_{cr}^{SC}$.
Note that $M_{cr}^{SC}$ becomes larger as we increase $\left\vert
\eta\right\vert $.

\item[$\left(  c\right)  $] Superluminal cases with $M_{cr}^{SC}=0$ where
$\eta<0$ and $0<n<2$. An example with $n=1$ is plotted in FIGs.
\ref{fig:etaBH:c} and \ref{fig:etaBH:d} for $\eta=-1,-10,-100,$ and $-1000$.
Opposite to the subluminal cases, the effects of rainbow gravity could
dramatically boost the luminosity for $M_{\ast\ast}\lesssim M\lesssim M_{\ast
}$ in this case if $\left\vert \eta\right\vert $ is large enough. For example,
the maximum value of the luminosity for $\eta=-1000$ is approximately $10^{3}$
times greater than that in the usual case and $10^{20}$ times greater for
$\eta=-10^{16}$, for which the energy scale of Lorentz-invariance violation is
assumed around $1$TeV. The rainbow gravity speeds up the process of
evaporation and hence the characteristic time scale for the black hole to
evaporate from $M_{\ast}$ to $M_{\ast\ast}$ becomes shorter than in the usual case.
\end{enumerate}

The luminosity for a Schwarzschild black hole has also been calculated in the
framework of rainbow gravity in \cite{IN-Esposito:2010pg}, where massless
bosons were considered and the authors used the MDR of the form of%
\begin{equation}
E^{2}-p^{2}\left(  1-\lambda p\right)  =m^{2}.
\end{equation}
By contrast, there are a number of differences between the calculations in our
paper and in \cite{IN-Esposito:2010pg}, which are as follows:

\begin{enumerate}
\item The geometric optics approximation have been used to calculate the
luminosity of a black hole in both papers. In such approximation, a
Schwarzschild black hole can be described as a black sphere of the radius $R$
and the temperature $T$. As a result, when the luminosity is calculated in the
framework of rainbow gravity, we have considered effects of rainbow gravity on
both $R$ and $T$. On the other hand, only effects of rainbow gravity on $T$
were considered in \cite{IN-Esposito:2010pg}. As discussed before, the
corrections to $R$ could play an important role in the late stage of the
evaporation process. This can be illustrated by FIG. \ref{fig:etaBH:d}, where
the corrections to $R$ dominate over those to $T$ and make the luminosity
smaller than in the usual case for small enough $M/m_{p}$, although the
corrections to $T$ tend to increase the luminosity.

\item In our paper, we used detailed balance condition to show that the
average number $n_{E,i}$ in the mode with $E$ and $i$ is%
\begin{equation}
n_{E,i}=n\left(  \frac{E}{T_{eff}}\right)  ,
\end{equation}
where $T_{eff}$ is given in eqn. $\left(  \ref{eq:Eff-Temp}\right)  $. In
contrast, the authors of \cite{IN-Esposito:2010pg} assumed an average behavior
for particles described by a unique average temperature, which is $T^{SC}$
obtained in section \ref{Sec:TBHRG}. Therefore in \cite{IN-Esposito:2010pg},
the average number $n_{E,i}$ was given by%
\begin{equation}
n_{E,i}=n\left(  \frac{E}{T^{SC}}\right)  .
\end{equation}

\item We have required the particles' energy $E\leq\min\left\{  m_{p}%
x_{cr},M-M_{cr}\right\}  $ while only $E\leq m_{p}x_{cr}$ was used in
\cite{IN-Esposito:2010pg}. The extra requirement $E\leq M-M^{cr}$ could
dramatically change behaviors of the evaporation process in the late stage. In
fact, for the usual case with $\eta=0,$ $x_{cr}=\infty,$ and $M_{cr}=0$, one
finds the luminosity of bosons without imposing $E\leq M-M^{cr}$ is $L\sim
M^{-2}$, which implies that the black hole would evaporate completely in
finite time and have a final explosion at $M=0$. However, if the requirement
$E\leq M-M^{cr}$ is imposed, the luminosity $L\sim M^{5}$ for small enough
$M$. Thus, it takes infinite time for the black hole to evaporate completely
and death of the black hole is much milder. In our paper, since a black hole
in rainbow gravity shares the similar the late-stage behaviors with that in
the usual case, the lifetime of the rainbow black hole is also infinite. In
contrast, the lifetime of a rainbow black hole in \cite{IN-Esposito:2010pg}
turned out to be finite.
\end{enumerate}

\subsection{4D Cylindrically Symmetric Black Hole}

For a 4D cylindrically symmetric black hole with $h_{ab}\left(  x\right)
dx^{a}dx^{b}=d\theta^{2}+\alpha^{2}dz^{2}$ in eqn. $\left(  \ref{eq:BHmetric}%
\right)  $, we have found that $\lambda$ in eqn. $\left(  \ref{eq:HJ-Lamda&W}%
\right)  $ is \cite{SF}%
\begin{equation}
\lambda=j^{2}\hbar^{2}+\frac{J_{z}^{2}}{\alpha^{2}},
\end{equation}
where $j$ is the angular momentum along $z$-axis and $J_{z}$ is a constant. To
count the number of modes of radiation, we assume the length of the black
string is $a$. Thus, the periodicity condition along $z$-axis gives%
\begin{equation}
J_{z}=\frac{2\pi k\hbar}{a}\text{ with }k\in Z.
\end{equation}
In the geometric optics approximation, eqn. $\left(  \ref{eq:HJ-Lamda&W}%
\right)  $ puts an upper bound on $\lambda$%
\begin{equation}
\lambda\leq\lambda_{\max}\equiv\frac{C\left(  r_{\min}^{2}\right)  }{B\left(
r_{\min}\right)  }\frac{f^{2}\left(  E/m_{p}\right)  }{g^{2}\left(
E/m_{p}\right)  }E^{2},
\end{equation}
where $\frac{C\left(  r^{2}\right)  }{B\left(  r\right)  }$ has a minimum at
$r_{\min}$. The luminosity per unit length $l$ is%
\begin{align}
l &  \equiv\frac{L}{a}=\frac{g_{s}}{a}%
{\displaystyle\sum\limits_{j,k}}
\int En\left(  \frac{E}{T_{0}}\frac{f\left(  E/m_{p}\right)  }{g\left(
E/m_{p}\right)  }\right)  \frac{dE}{2\pi\hbar}\nonumber\\
&  =\frac{\alpha g_{s}}{4\pi^{2}\hbar^{3}}\int d\left(  j\hbar\right)
d\left(  \frac{J_{z}}{\alpha}\right)  \int En\left(  \frac{E}{T_{0}}%
\frac{f\left(  E/m_{p}\right)  }{g\left(  E/m_{p}\right)  }\right)
dE\nonumber\\
&  =\frac{\alpha g_{s}}{2\pi\hbar^{3}}\frac{C\left(  r_{\min}^{2}\right)
}{B\left(  r_{\min}\right)  }\int_{0}^{E_{\max}}\frac{f^{2}\left(
E/m_{p}\right)  }{g^{2}\left(  E/m_{p}\right)  }E^{3}n\left(  \frac{E}{T_{0}%
}\frac{f\left(  E/m_{p}\right)  }{g\left(  E/m_{p}\right)  }\right)
dE\nonumber\\
&  =\frac{\alpha g_{s}T_{0}^{4}}{2\pi m_{p}^{4}}\frac{C\left(  r_{\min}%
^{2}\right)  }{B\left(  r_{\min}\right)  m_{p}^{2}}\int_{0}^{u_{\max}}h\left(
\frac{uT_{0}}{m_{p}}\right)  \left[  h\left(  \frac{uT_{0}}{m_{p}}\right)
+\left(  \frac{uT_{0}}{m_{p}}\right)  h^{\prime}\left(  \frac{uT_{0}}{m_{p}%
}\right)  \right]  u^{3}n\left(  u\right)  du,
\end{align}
where $u=\frac{E}{T_{0}}\frac{f\left(  E/m_{p}\right)  }{g\left(
E/m_{p}\right)  }$.

For a static uncharged black string $\left(  \ref{eq:BSMetric}\right)  $, one
has%
\[
\text{ }T_{0}=\frac{3\alpha m_{p}}{4\pi}b^{1/3},\text{ }r_{\min}=\infty,\text{
and }\frac{C\left(  r_{\min}^{2}\right)  }{B\left(  r_{\min}\right)  }%
=\alpha^{-2},
\]
where $b=4M.$ Since the length of the black string is infinite, one only has
$E\leq E_{\max}\equiv m_{p}x_{cr}$. The luminosity per unit length for a
static uncharged black string is%
\begin{align}
l &  =\alpha\frac{81m_{p}^{2}\alpha^{2}g_{s}}{512\pi^{5}}b^{\frac{4}{3}}%
\int_{0}^{\frac{4\pi y_{cr}}{3\alpha m_{p}b^{1/3}}}h\left(  \frac{3\alpha
m_{p}}{4\pi}b^{1/3}u\right)  \nonumber\\
&  \left[  h\left(  \frac{3\alpha m_{p}}{4\pi}b^{1/3}u\right)  +\left(
\frac{3\alpha m_{p}}{4\pi}b^{1/3}u\right)  h^{\prime}\left(  \frac{3\alpha
m_{p}}{4\pi}b^{1/3}u\right)  \right]  u^{3}n\left(  u\right)
du.\label{eq:L-BS-h}%
\end{align}
In the cases with $M_{cr}^{BS}=0$, if $T_{0}\ll m_{p}\left(  M\ll1\right)  $,
for the emission of $n_{s}$ species of massless particles of spin $s,$ we have%
\begin{align}
l &  \approx\alpha\frac{81m_{p}^{2}\alpha^{2}}{512\pi^{5}}b^{\frac{4}{3}%
}\left\{  \left(  n_{0}+2n_{1}+n_{1/2}\right)  \right.  \nonumber\\
&  \left.  -\frac{3\left(  n+1\right)  !\left(  n+2\right)  }{\pi^{2}}\left(
\frac{3\alpha m_{p}}{4\pi}b^{1/3}u\right)  ^{n}\eta\left[  Li_{n+2}\left(
1\right)  \left(  n_{0}+2n_{1}\right)  +\frac{2-2^{-n}}{n+2}\zeta\left(
n+2\right)  n_{1/2}\right]  \right\}  ,\label{eq:L-BS}%
\end{align}
where $Li_{s}\left(  z\right)  $ is the polylogarithm of order $s$ and
argument $z$ and $\zeta\left(  z\right)  $ is the zeta function. From
eqn.$\left(  \ref{eq:L-BS}\right)  $, we find that $\frac{dM}{dt}\equiv l\sim
M^{\frac{4}{3}}$ for $M\ll1.$Consequently, just like the usual case, we have
for the rainbow black string that $M\sim t^{-3}$, which means that its
lifetime is infinite.

\begin{figure}[tb]
\begin{center}
\subfigure[{~ Subluminal cases and superluminal cases with and without remnant}]{
\includegraphics[width=0.48\textwidth]{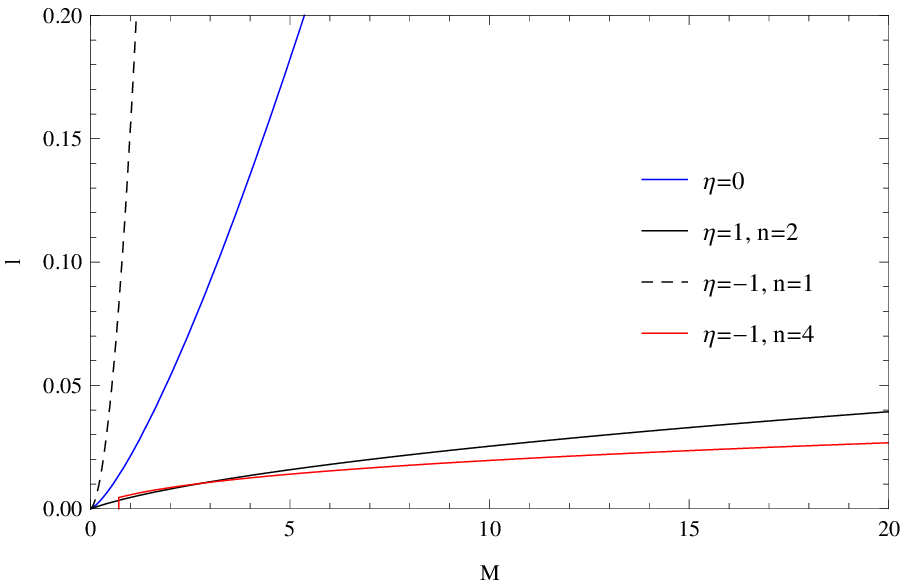}\label{fig:LBS:a}}
\subfigure[{~Subluminal cases for various values of $\eta$}]{
\includegraphics[width=0.48\textwidth]{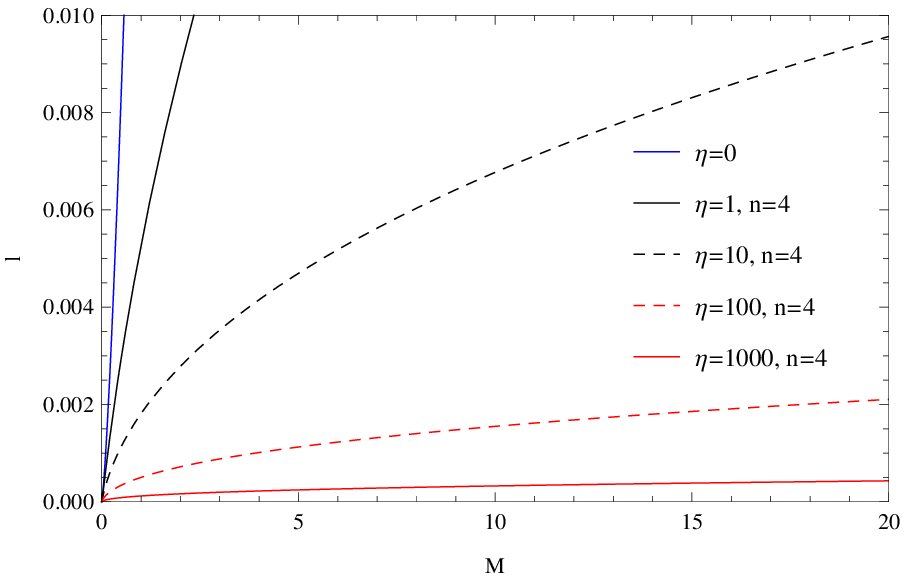}\label{fig:LBS:b}}
\end{center}
\caption{Plots of the luminosity per unit length $l$ of radiation of one
species of bosons against the mass per unit length $M$ for a static uncharged
black string. }%
\label{fig:LBS}%
\end{figure}

In the following, we will compute the asymptotic value of $l$ when $T_{0}\gg
m_{p}\left(  M\gg1\right)  $. In the usual case, the luminosity per unit
length $l\propto M^{\frac{4}{3}}$. For other cases, the results show as follows:

\begin{enumerate}
\item[$\left(  a\right)  $] Subluminal cases where $\eta>0$ and $M_{cr}%
^{BS}=0$. Using
\begin{equation}
x=yh\left(  y\right)  \sim\eta^{-1/n}-\frac{\eta^{-3/n}}{ny^{2}},
\end{equation}
in eqn. $\left(  \ref{eq:L-BS-h}\right)  $, we have for the luminosity per
unit length%
\begin{equation}
l\sim\frac{\eta^{-\frac{4}{n}}g_{s}}{\pi nm_{p}^{2}\alpha}\int_{\frac{4\pi
}{3\alpha m_{p}\left(  4M\right)  ^{1/3}}}^{\infty}\frac{n\left(  \tilde
{u}\right)  }{\tilde{u}}d\tilde{u}.
\end{equation}
Thus, the luminosity per unit length of radiation of one species of bosons is%
\begin{equation}
l\sim\frac{3\eta^{-\frac{4}{n}}g_{s}}{4n\pi^{2}m_{p}}\left(  4M\right)
^{1/3},
\end{equation}
and that of radiation of one species of fermions is%
\begin{equation}
l\sim\frac{\eta^{-\frac{4}{n}}g_{s}}{3n\pi m_{p}^{2}\alpha}\ln M.
\end{equation}
The luminosity per unit length $l$ is lower than that in the usual case for
large $M$.\ In addition, the radiation of bosons dominates the evaporation
process for $M\gg1$. It is evident that $l$ becomes smaller as $\eta$ is
increased. An example of one species of bosons with $\eta=1$ and $n=2$ is
plotted as a black line in FIG. \ref{fig:LBS:a}. Additionally, we plot $l$
against $M$ for the examples with $n=4$ in FIG. \ref{fig:LBS:b} for
$\eta=1,10,100,$ and $1000$.

\item[$\left(  b\right)  $] Superluminal cases with the non-vanishing remnant
mass $M_{cr}^{BS}$ where $\eta<0$ and $n\geq2$. In this case, one has for
radiation of one species of bosons%
\begin{equation}
l\sim\frac{3g_{s}}{8\pi^{2}}\frac{\left(  4M\right)  ^{1/3}}{m_{p}}\int
_{0}^{y_{cr}}h\left(  y\right)  \left[  h\left(  y\right)  +yh^{\prime}\left(
y\right)  \right]  y^{2}dy,
\end{equation}
and radiation of one species of fermions%
\begin{equation}
l\sim\frac{g_{s}}{2\pi\alpha m_{p}^{2}}\int_{0}^{y_{cr}}h\left(  y\right)
\left[  h\left(  y\right)  +yh^{\prime}\left(  y\right)  \right]
y^{3}dy\text{.}%
\end{equation}
Similar to the subluminal cases, $l$ could become much lower than that in the
usual case for large enough $M$ and the radiation of bosons dominates the
evaporation process for $M\gg1$. An example of one species of bosons with
$\eta=-1$ and $n=4$ is plotted as a red solid line in FIG. \ref{fig:LBS:a}.

\item[$\left(  c\right)  $] Superluminal cases with $M_{cr}^{BS}=0$ where
$\eta<0$ and $0<n<2$. Using
\begin{equation}
x=yh\left(  y\right)  \sim\left(  -\eta\right)  ^{\frac{1}{2-n}}y^{\frac
{2}{2-n}}%
\end{equation}
in eqn. $\left(  \ref{eq:L-BS-h}\right)  $, we find%
\begin{equation}
l\sim\frac{\left(  -\eta\right)  ^{\frac{2}{2-n}}g_{s}}{2-n}\frac{81m_{p}%
^{2}\alpha^{3}}{256\pi^{5}}\left(  \frac{3\alpha m_{p}}{4\pi}\right)
^{\frac{n}{2-n}}\left(  4M\right)  ^{\frac{8-3n}{3\left(  2-n\right)  }}%
\int_{0}^{\infty}u^{2}n\left(  u\right)  u^{\frac{2+n}{2-n}}du.
\end{equation}
Since $\frac{8-3n}{3\left(  2-n\right)  }>\frac{4}{3}$ for $0<n<2\,,$ $l$
could become much larger than that in the usual case for large enough $M$. The
larger $\left\vert \eta\right\vert $ is, the faster the black string
evaporates for $M\gg1$. An example of one species of bosons with $\eta=-1$ and
$n=1$ is plotted as a black dashed line in FIG. \ref{fig:LBS:a}.
\end{enumerate}

\section{Conclusion}

\label{Sec:Con}

In this paper, we have analyzed the effects of rainbow gravity on the
temperatures, entropies and luminosities of black holes. Using the
Hamilton-Jacobi method for scalars, spin $1/2$ fermions and vector bosons, we
first obtained the effective temperature $T_{eff}$ of a black hole, which
depends on the energy $E$ of emitted particles. By relating the momentum $p$
of particles to the event horizon radius $r_{h}$ of the black hole, the
temperatures of a rainbow Schwarzschild black hole and a rainbow static
uncharged black string were calculated. Focusing on the AC dispersion relation
with $f\left(  x\right)  =1$ and $g\left(  x\right)  =\sqrt{1-\eta x^{n}}$, we
computed their minimum masses $M_{cr}$ and final temperatures $T_{cr}\ $for
different values of $\eta$ and $n$. All the results were listed in TABLE
\ref{tab:one}. In addition, a non-vanishing minimum mass indicates the
existence of the black hole's remnant, which could shed light on the
\textquotedblleft information paradox\textquotedblright. The entropies were
also studied in section \ref{Sec:TBHRG}.

In section \ref{Sec:LBHRG}, we used the geometric optics approximation to
compute luminosities of a 2D black hole, a Schwarzschild one and a static
uncharged black string in the framework of rainbow gravity. It was found that
the luminosity of the rainbow Schwarzschild black hole with the mass $M$
deviates from that in the usual case only when $M_{\ast\ast}\lesssim M\lesssim
M_{\ast}$ for $M_{cr}^{SC}=0$ or $M_{cr}^{SC}\leq M\lesssim M_{\ast}$ for a
non-vanishing $M_{cr}^{SC}$, where $M_{\ast}\sim\left\vert \eta\right\vert
^{1/n}m_{p}$ and $M_{\ast\ast}\sim\left\vert \eta\right\vert ^{-1/n}m_{p}$. In
the subluminal cases where $\eta>0$ and $M_{cr}^{SC}=0$, FIG.
\ref{fig:etaBH:a} shows that the effects of rainbow gravity could
significantly suppress the luminosity for large $\eta$. In the superluminal
cases with $M_{cr}^{SC}=0$ where $\eta<0$ and $0<n<2$, FIG. \ref{fig:etaBH:c}
shows that the luminosity could be significantly boosted for large $\left\vert
\eta\right\vert $. Similar results for the rainbow static uncharged black
string with the mass per unit length $M\gg1$ were also obtained for the
subluminal cases and superluminal cases with $M_{cr}^{BS}=0$.

If the energy scale of Lorentz-invariance violation is $\Lambda$, the
naturalness in effective field theories implies that $\left\vert
\eta\right\vert \sim\left(  \frac{\Lambda}{m_{p}}\right)  ^{n}$. \ For a
Schwarzschild black hole, the effects of rainbow gravity start to play an
important role when the mass of the black hole $M\lesssim$ $\left\vert
\eta\right\vert ^{1/n}m_{p}\sim\Lambda$. If Lorentz invariance is violated by
quantum gravity, the natural scale $\Lambda\sim m_{p}$. Currently, there are
no experimental evidence that Lorentz symmetry is violated in nature, which
might suggest that $1$TeV$\lesssim\Lambda\lesssim m_{p}$. The experimental and
observational constraints on Lorentz-invariance violation are reviewed in
\cite{CON-Mattingly:2005re,CON-Liberati:2013xla}. As a result, the rainbow
gravity plays a negligible role for stellar or galactic supermassive black
holes. However, the possible production of TeV-scale black holes at the LHC or
ultra-high-energy cosmic ray collisions is predicted by low-scale quantum
gravity\cite{CON-Landsberg:2002sa}. Another possible source of small black
holes is primordial black holes, which are created by primordial density
fluctuations in the early universe and evaporate for enough long time. Future
studies investigating the implications of our results on the rich
phenomenology of these small black holes would be interesting.

\begin{acknowledgments}
We are grateful to Houwen Wu and Zheng Sun for useful discussions. This work
is supported in part by NSFC (Grant No. 11005016, 11175039 and 11375121) and
the Fundamental Research Funds for the Central Universities.
\end{acknowledgments}

\end{document}